\documentstyle[11pt,aaspp4,psfig]{article}

\begin{document}        

\title{Hybrid Thermal-Nonthermal Synchrotron Emission from Hot Accretion Flows}
\author{Feryal \"Ozel\altaffilmark{1}, Dimitrios Psaltis, and Ramesh Narayan}
\affil{Harvard-Smithsonian Center for Astrophysics, 60 Garden 
Street, Cambridge, MA 02138; \\
fozel,dpsaltis,rnarayan@cfa.harvard.edu}
\altaffiltext{1}{Physics Department, Harvard University}

\begin{abstract}
  
  We investigate the effect of a hybrid electron population,
  consisting of both thermal and non-thermal particles, on the
  synchrotron spectrum, image size, and image shape of a hot accretion
  flow onto a supermassive black hole. We find two universal features
  in the emitted synchrotron spectrum: (i) a prominent shoulder at low
  $(\lesssim 10^{11}~\rm{Hz})$ frequencies that is weakly dependent on
  the shape of the electron energy distribution, and (ii) an extended
  tail of emission at high $(\gtrsim 10^{13}~\rm{Hz})$ frequencies
  whose spectral slope depends on the slope of the power-law energy
  distribution of the electrons. In the low-frequency shoulder, the
  luminosity can be up to two orders of magnitude greater than with a
  purely thermal plasma even if only a small fraction $(< 1\%)$ of the
  steady-state electron energy is in the non-thermal electrons.  We
  apply the hybrid model to the Galactic center source, Sgr~A$^*$. The
  observed radio and IR spectra imply that at most $1\%$ of the
  steady-state electron energy is present in a power-law tail in this
  source. This corresponds to no more than $10\%$ of the electron
  energy injected into the non-thermal electrons and hence $90 \%$
  into the thermal electrons. We show that such a hybrid distribution
  can be sustained in the flow because thermalization via Coulomb
  collisions and synchrotron self-absorption are both inefficient.
  
  The presence of non-thermal electrons enlarges the size of the radio
  image at low frequencies and alters the frequency dependence of the
  brightness temperature. A purely thermal electron distributions
  produces a sharp-edged image while a hybrid distribution causes
  strong limb brightening. These effects can be seen up to frequencies
  $\sim 10^{11}$~Hz and are accessible to radio interferometers.

\end{abstract}


\centerline{To appear in {\it The Astrophysical Journal}}

\section{Introduction}

The mechanisms of particle heating and acceleration, and the emission
spectra from the resulting particle energy distributions, are of great
importance in the theory of collisionless hot accretion flows onto
compact objects.  Discussions in the literature have focused on the
physics of electron heating and acceleration (Begelman \& Chiueh 1988;
Bisnovatyi-Kogan \& Lovelace 1997; Quataert \& Gruzinov 1999; Gruzinov
\& Quataert 1999, Medvedev 2000), the efficiency of particle
thermalization (Ghisellini, Guilbert, \& Svensson 1988; Ghisellini,
Haardt, \& Fabian 1993; Mahadevan \& Quataert 1997; Ghisellini,
Haardt, \& Svensson 1998; Nayakshin \& Melia 1998), and the generation
of hybrid thermal-nonthermal electron energy distributions in these
plasmas (see, e.g., Coppi 1999 and references therein).  Despite a
substantial amount of work, many issues remain unresolved, primarily
because of our incomplete understanding of physical processes such as
magnetic reconnection, MHD turbulence, and collective plasma modes.

These questions are especially relevant for an optically-thin
advection-dominated accretion flow (ADAF). An ADAF is an example of a
hot, rarefied, magnetic plasmas with low radiative efficiency
(Ichimaru 1977; Narayan \& Yi 1994, 1995b; Abramowicz et al. 1995; see
Narayan, Mahadevan, \& Quataert 1998b and Kato, Fukue, \& Mineshige
1998 for reviews). A basic property of the nearly collisionless plasma
in an ADAF is that the Coulomb coupling between the electrons and ions
is weak so that energy transfer from the ions to the electrons is
inefficient. In addition, it is commonly assumed that the viscously
generated energy primarily heats the heavier species, the ions, and
that the electrons retain a thermal distribution throughout the flow
(e.g., Narayan \& Yi 1995b; Mahadevan 1997).

There are, however, processes such as MHD turbulence, pair production
(e.g., through pion decay), and electron-proton coupling which can
both heat the electrons and generate non-thermal distributions.
Quataert and Gruzinov (1999; see also Gruzinov \& Quataert 1999)
considered two processes specific to MHD turbulence that accelerate
particles in magnetic collisionless plasmas: Landau damping by
electric fields parallel to the local magnetic field and transit-time
damping by time-varying magnetic fields. They found that the
assumption of negligible electron heating/acceleration is valid only
for weak magnetic fields, i.e., when the ratio of the gas pressure to
total pressure $\beta_{\rm{ADAF}}$ is larger than a critical value
$\beta_{\rm{crit}}$. The value of $\beta_{\rm{crit}}$ is very
uncertain and is around 0.9. For $\beta_{\rm{ADAF}} \gtrsim 0.9$,
turbulence primarily accelerates the protons, while for stronger
magnetic fields, the results are inconclusive. Shocks and pion decay
can also lead to non-thermal electrons in the accretion flow.

Several processes, such as Coulomb collisions and synchrotron
self-absorption, can potentially lead to thermalization of particles
in accretion flows (Mahadevan \& Quataert 1997; Ghisellini et al.
1998).  Mahadevan \& Quataert (1997) showed that Coulomb collisions
are ineffective in thermalizing the electrons in an ADAF. However,
they argued that, for sufficiently high mass-accretion rates, the
electrons in the plasma can be thermalized by synchrotron
self-absorption. Nayakshin and Melia (1998) showed that considerable
deviations from a Maxwell-Boltzmann distribution can be sustained in a
plasma with low source compactness when Coulomb collisions,
Comptonization, and pair processes are taken into account.

In view of the difficulty of calculating the heating, cooling and
thermalization of particles from first principles, many authors have
restricted their models to either purely Maxwellian or purely
non-thermal (extended power laws or monoenergetic) electron
distributions. Only recently have there been attempts towards
explaining spectra of accreting black holes with models including
hybrid thermal/non-thermal distributions of electrons. Some models
have physically motivated distributions, such as non-thermal electrons
produced by decaying pions (Mahadevan 1998), whereas others invoke
more {\it ad hoc} distributions to fit the data (e.g., Beckert \&
Duschl 1997; Falcke \& Biermann 1999). The reverse process of trying
to constrain the energy distributions of particles in accreting
plasmas by comparing models to data, however, faces issues of
uniqueness which can be addressed only by a more comprehensive study
of hybrid models.

In this paper, we consider generalized electron distributions
consisting of a dominant Maxwellian plus a small non-thermal
power-law component of varying slope and energy content, and study the
synchrotron emission from the resulting hybrid plasmas. We identify
the characteristic signatures of the non-thermal electrons on the
emitted radio synchrotron spectrum of an accretion
flow and on its image as observed with a radio telescope. We also discuss to
what extent the observed effects could be used to determine the
details of the underlying non-thermal electron distribution. This work
is relevant for interpreting observations of low luminosity AGNs such
as the Galactic Center source, Sagittarius~A$^*$ (Sgr~A$^*$).

In $\S 2$ we review the basic properties of ADAFs, hybrid plasmas,
synchrotron radiation, and radiative transfer. In $\S 3$ we present a
series of models of supermassive black holes with low accretion rates.
We apply these models to Sgr~A$^*$ and derive constraints on the
fraction of the electron energy that can be present in a non-thermal
form.  In $\S 4$, we study the correspondence between the energy
distribution of the electrons and the resulting spectrum. In $\S5$ we
study the energetics of a hybrid flow and calculate the heating,
cooling and thermalization rates of the non-thermal electrons. We summarize
our conclusions in $\S 6$. We present in an Appendix
approximate analytic expressions for the contribution of a non-thermal
particle population to the synchrotron spectrum of an accretion flow.

\section{Formalism}

\subsection{Advection-Dominated Flows}

We begin by reviewing some of the basic properties of the
optically-thin branch of ADAFs. ADAFs are quasi-spherical, hot,
magnetic accretion flows in which the accreting plasma is too rarefied
to cool efficiently by radiative processes.  The viscously dissipated
energy is therefore advected into the black hole or other compact
object at the center (see Narayan et al. 1998b and Kato et al. 1998 for
reviews).  In the limit where the fraction of the viscous energy
advected inward is independent of radius, a self-similar analytic
solution for the thermodynamic quantities of the accreting gas can be
obtained (Narayan \& Yi 1994, 1995b).  We make use of this solution in
Appendix A. For the numerical calculations presented in the rest of
the paper, we use more accurate global solutions to obtain the run of
electron temperature and density with radius. These solutions are
calculated by the methods described in Narayan, Kato, \& Honma (1997b),
Chen, Abramowicz, \& Lasota (1997), and Popham \& Gammie (1998).  The
magnetic field strengths are obtained by assuming that the ratio of
gas pressure to total pressure (sum of gas and magnetic pressure) is
equal to a specified value $\beta_{\rm{ADAF}}$.

Since the focus of this paper is on massive black holes in galactic
nuclei with low mass-accretion rates, with specific applications to
the black hole in our own Galactic nucleus, Sgr~A$^*$, we scale masses
in units of $10^{6}$ solar mass, i.e., $M \equiv m_{6} 10^{6}
M_{\odot},$ and radii in units of the Schwarzschild radius, i.e., $R
\equiv r R_{Sch}$, where
\begin{equation}
R_{Sch} = \frac{2 G M}{c^2} = 2.95 \times 10^{11} m_6 ~ {\rm cm.}
\end{equation}  
We scale the mass accretion rate in units of $10^{-3}
\dot{M}_{\rm{Edd}}$, i.e., $\dot{M} \equiv \dot{m}_{-3} 10^{-3}
\dot{M}_{\rm{Edd}},$ where the Eddington mass accretion rate is
\begin{equation}
\dot{M}_{\rm{Edd}} = \frac{L_{\rm{Edd}}}{\eta_{\rm{eff}} c^{2}} 
= 1.39 \times 10^{24} 
\left(\frac{\eta_{\rm{eff}}}{0.1}\right)^{-1} m_{6} ~ {\rm g ~ s^{-1}}. 
\end{equation}
In defining the Eddington rate, we assume a standard radiative
efficiency of $\eta_{\rm{eff}} = 0.1$. (This is purely for the
purposes of the definition; the actual radiative efficiency can be
very different from $0.1$).

For the calculations presented here we use ADAF models in which the
viscosity parameter is set to $\alpha = 0.1$, the equipartition
parameter to $\beta_{\rm{ADAF}} = 0.968$, and the ratio of viscous electron
heating to proton heating to $\delta = 10^{-3}$. Note that
$\beta_{\rm{ADAF}}$ differs from the usual plasma parameter $\beta$, which
is the ratio of the gas pressure to the magnetic pressure; the value
of $\beta_{\rm{ADAF}}=0.968$ assumed in our ADAF models corresponds to a
plasma $\beta=10$ (Quataert \& Narayan 1999).

\subsection{Hybrid Populations}

We assume that a large fraction of the electrons in the plasma are in a
thermal distribution with temperature $T$ and that the rest of the
electrons are in a non-thermal distribution, usually with a power-law
form. We denote the number density of electrons in the thermal
population by $N_{\rm{th}}$ and in the non-thermal population by
$N_{\rm{pl}}$. We denote the emissivities of the thermal and power-law
electron populations by $j_{\rm{th}}$ and $j_{\rm{pl}}$ respectively,
and the corresponding absorption coefficients by $\alpha_{\rm{th}}$
and $\alpha_{\rm{pl}}$.

For the thermal electron population, we use 
the relativistic Maxwell-Boltzmann distribution given by
\begin{equation}
 n_{\rm{th}} (\gamma) = N_{\rm{th}} \gamma^{2} \beta \rm{exp}
(-\gamma / \theta_{e}) / [\theta_{e} K_{2} (1/\theta_{e})], 
\end{equation}
where $\gamma$ is the electron Lorentz factor, $\beta$ is the
relativistic electron velocity, and $\theta_e \equiv k T / m_e c^2$ is
the dimensionless electron temperature.  The modified Bessel function
of second order $K_2(1/\theta_e)$ arises from the normalization of the
Maxwellian.  Similarly, for the non-thermal electron population we use
a power-law distribution extending from $\gamma = 1$ to infinity,
\begin{equation}
n_{\rm{pl}}(\gamma) = N_{\rm{pl}} (p-1) \gamma^{-p}.
\end{equation}

The number density of thermal electrons $N_{\rm{th}}$ is a function of
the flow radius and is determined by the global ADAF solutions. We
determine $N_{\rm{pl}}$ at each radius by assuming that the
steady-state energy in the non-thermal distribution is equal to a
fraction $\eta$ of the energy in the thermal distribution, with $\eta$
constant in radius. Although the calculations presented in $\S 3$ are
all carried out with this assumption, generalizations to
radially-dependent $\eta(r)$ as well as a discussion of the energetics
of such a flow will be presented in $\S 5$. Note that we
implicitly assume that the non-thermal electron population does not
affect the dynamics or the thermal properties of the flow; this will
be justified in $\S 5$. 

The energy density of a
Maxwell-Boltzmann distribution of electrons at temperature $\theta_e$
was derived by Chandrasekhar $(1939,\rm{eq}. ~ [236])$ to be
\begin{equation}
u = a\left(\theta_e \right) N_{\rm{th}} m_e c^2 \theta_e, 
\end{equation}
where 
\begin{equation}
a\left(\theta_e\right) \equiv \frac{1}{\theta_e}\left[\frac{3 K_3(1/\theta_e)
    + K_1(1/\theta_e)}{4 K_2(1/\theta_e)} - 1\right]
\end{equation}
is a coefficient that varies from 3/2 for a non-relativistic electron
gas to 3 for fully relativistic electrons, and $K_n$ are the modified
Bessel functions of the n$\it{th}$ order. For the present purposes, we
use a simplified expression for $a(\theta_{e})$ which has an error of
less than $2\%$ at all temperatures (Gammie \& Popham 1998):
\begin{equation}
a(\theta_e) = \frac{6 + 15 \theta_e}{4+5\theta_e}.
\end{equation}
The number density of the non-thermal distribution is then
\begin{equation}
  \label{eq:normalization}
 N_{\rm{pl}} = \eta a(\theta_e) \theta_e (p-2) N_{\rm{th}}.
\end{equation}
This normalization of the power law population typically corresponds to 
\begin{equation}
  \label{eq:number}
\frac{N_{\rm{pl}}}{N_{\rm{th}}} \sim (0.1-10)\eta, 
\end{equation}
depending on the electron temperature and the power law index. 

The distributions considered above correspond to the steady state that
results from the competition between heating/acceleration and cooling
by radiation. We discuss in some detail in $\S 5$ the energy equations
for the thermal and non-thermal electron populations.  Here we simply
note that the synchrotron cooling timescale of electrons moving with a
Lorentz factor $\gamma$ scales as $\gamma^{-2}$, and hence electrons
in the high energy tail of a power law distribution cool most
rapidly. As a result, if electrons are injected with a power-law
distribution with index $s$ $[n(\gamma) \propto \gamma^{-s}]$ and cool
only by synchrotron emission, the synchrotron cooling causes the power
law index of the steady state distribution to be $p = s+1$ above a
certain $\gamma_b$, called the cooling break, thus causing the steady
state distribution to fall off more steeply at higher electron
energies.  Mahadevan \& Quataert (1997) calculated $\gamma_b$ in an
ADAF by comparing the cooling timescale to the inflow timescale and
found that at sufficiently high accretion rates, the break occurs at a
very low Lorentz factor, $\gamma_b \sim 1.5$. The Lorentz factors of
interest to us are invariably larger than $\gamma_b$. Therefore, the
values of $p$ we consider below in $\S 3$ and $\S 4$ are always equal
to $s+1$, so that the injected energy distribution $\gamma^{-s}$ is
harder by one power of $\gamma$ than the steady state energy
distribution, $\gamma^{-p}$.

Corresponding to $\eta$, we can define another quantity
$\eta_{\rm{inj}}$ that measures the fraction of electron energy
$\it{injected}$ into a power law distribution. If $s < 2$, then
$\eta_{\rm{inj}}$ can be significantly greater than $\eta$. If we 
assume a distribution from $\gamma_{\rm{min}} = 1$ to some 
$\gamma_{\rm{max}}$, $\eta_{\rm{inj}}$ is greater than $\eta$ by a
factor $\sim \gamma_{\rm{max}}^{3-p}$. However, it is thought that the
acceleration mechanisms typically encountered in astrophysics, such as
shock acceleration or acceleration via MHD turbulence, inject energy
into particles with $s > 2$ such that the steady state distribution
has  $p > 3$. In this case there is little dependence on
$\gamma_{\rm{max}}$, and
\begin{equation}
  \label{eq:etaeff}
  \frac{\eta_{\rm{inj}}}{\eta} \simeq \frac{p-2}{p-3}, 
\end{equation}
which is not very different from unity. Although we expect $s > 2$ and
$p>3$ for most systems, for completeness we consider models in the
range $2 < p < 4$.

\subsection{Synchrotron Emissivity}

The synchrotron emissivity of a relativistic electron moving with a
Lorentz factor $\gamma$ in a magnetic field of strength $B$ is given
by (Rybicki \& Lightman 1979)
\begin{equation}
  \label{eq:sngl}
j_{\nu}(\gamma, \theta) = \frac{\sqrt{3} e^{2}}{2 c} \nu_{b} \sin \theta F(x). 
\end{equation} 
Here, $\nu_b \equiv e B / 2 \pi m_e c$ is the non-relativistic
cyclotron frequency, $\theta$ is the angle between the direction of
the magnetic field and the velocity of the electron, and
\begin{equation}
  \label{eq:fx}
F(x) \equiv x \int\limits_x^\infty K_{5/3}(t) dt, 
\end{equation}
with $K_{5/3}$ the modified Bessel function of order 5/3, $x \equiv
\nu/\nu_c$, and $\nu_c \equiv \frac{3}{2} \gamma^2 \nu_b \sin \theta$.
For a thermal distribution of electrons, the total emissivity for a
given angle $\theta$ is obtained by integrating equation
(\ref{eq:sngl}) over the Maxwellian distribution (Pacholczyk 1970),
\begin{equation}
  \label{eq:ixm}
j_{\nu,\rm{th}}(\theta) = \frac{N_{\rm{th}} e^2}{\sqrt{3} c K_2(1/\theta_e)} 
  \nu I\left(\frac{x_M}{\sin \theta}\right), 
\end{equation}
where 
\begin{equation}
  \label{eq:xm}
  x_M \equiv \frac{2 \nu}{3 \nu_b \theta_e^2}
\end{equation}
and
\begin{equation}
  \label{eq:ixmdef}
I(x_M) \equiv \frac{1}{x_M} \int\limits_0^\infty z^2 
\exp (-z) F(x_M/z^2) dz.
\end{equation}
The limiting behaviour of $I(x_M)$ for small and large $x_M$ was
derived by Pacholczyk $(1970)$ and Petrosian $(1981)$, respectively.
Mahadevan, Narayan, \& Yi (1996, hereafter MNY96) integrated equation
(\ref{eq:ixm}) over the angle $\theta$ for an isotropic distribution
of electron velocities and provided a fitting function for the
emissivity in the ultrarelativistic to mildly relativistic regimes:
\begin{equation}
  \label{eq:jtot}
  j_{\nu,\rm{th}} = \frac{N_{\rm{th}} e^2}{\sqrt{3} c K_2(1/\theta_e)} \nu M(x_M), 
\end{equation}
with $M(x_M)$ given by
\begin{equation}
  \label{eq:mxm}
M(x_{M}) = \frac{4.0505~a}{x_{M}^{1/6}} \left(1 + \frac{0.40~b}{x_{M}^{1/4}} +  
      \frac{0.5316~c}{x_{M}^{1/2}}\right) \exp(-1.8896 x_{M}^{1/3}). 
\end{equation}
The best fit values of the coefficients $a, b$, and $c$ for different
temperatures are given in MNY96. The coefficients tend to unity in the
ultrarelativistic limit. Finally, the synchrotron absorption
coefficient $\alpha_{\rm{th}}$ is related to the emissivity via
Kirchoff's law,
\begin{equation}
 \alpha_{\nu,\rm{th}} = j_{\nu,\rm{th}}/B_{\nu}(T), 
\end{equation}
where $B_{\nu}(T)$ is the black body source function.

For the total emissivity of electrons in a power law distribution, we
use the expression given in Rybicki \& Lightman (1979) and average
over angles,
\begin{equation}
  \label{eq:jpl2}
 j_{\nu,\rm{pl}} =  C^j_{\rm{pl}} \eta \frac{e^2 N_{\rm{th}}}{c} 
a(\theta_e) \theta_e \nu_b \left(\frac{\nu}{\nu_b}\right)^{(1-p)/2}, 
\end{equation}
where $N_{\rm{pl}}$ is defined in terms of $N_{\rm{th}}$ as above and 
\begin{equation}
  \label{eq:cjpl}
C^j_{\rm{pl}} = \frac{\sqrt{\pi} 3^{p/2}}{4} \frac{(p-1) (p-2)}{(p+1)} 
 \frac{\Gamma(\frac{p}{4}+\frac{19}{12})
 \Gamma(\frac{p}{4}-\frac{1}{12}) \Gamma(\frac{p}{4}+\frac{5}{4})}
{\Gamma(\frac{p}{4}+\frac{7}{4})}.
\end{equation}
The corresponding absorption coefficient is
\begin{equation}
  \label{eq:apl1}
    \alpha_{\nu,\rm{pl}} = C^\alpha_{\rm{pl}} \eta \frac{e^2 
     N_{\rm{th}}}{c} a(\theta_e) \theta_e 
      \left(\frac{\nu_b}{\nu}\right)^{(p+3)/2} \nu^{-1}, 
\end{equation}
with 
\begin{equation}
  \label{eq:capl}
C^\alpha_{\rm{pl}} =  \frac{\sqrt{3 \pi} 3^{p/2}}{8}
\frac{(p-1) (p-2)}{m_e} \frac{\Gamma(\frac{3p+2}{12})
  \Gamma(\frac{3p+22}{12}) \Gamma(\frac{6+p}{4})} {\Gamma(\frac{8+p}{4})}.
\end{equation}

\subsection{Radiative Transfer and Numerical Methods}

The equation of radiative transfer for a time-independent, spherically
symmetric flow is (e.g., Mihalas 1978)
\begin{equation}
  \label{eq:radtr1}
\mu  (\partial/\partial r) + r^{-1} (1 - \mu^2)
 (\partial/\partial\mu)] I(r, \mu, \nu) = 
j(r,\nu) - \alpha(r, \nu) I(r, \mu, \nu),
\end{equation}
where $\mu \equiv \cos\theta = (dz/ds)$ is the cosine of the angle
between the ray and the radial direction, $r$ is the radial
coordinate, $\nu$ is the frequency, and $j$ and $\alpha$ are the
emission and absorption coefficients defined above. One can simplify
this equation by taking plane parallel rays of varying impact
parameters (perpendicular distances of rays to the central line of sight)
through the flow and solving the equation along these rays (Mihalas
1978). The equation then becomes
\begin{equation}
  \label{eq:radtr2}
\pm \frac{\partial{I^{\pm}_\nu}}{\partial{s}} = 
  j_\nu - \alpha_\nu I^{\pm}_\nu , 
\end{equation}
where now $s$ is the line element along the ray and the coefficients
$+1$ and $-1$ correspond to radiation coming towards and going away
from an external observer, respectively. In our problem, $j_\nu =
j_{\rm{th}} + j_{\rm{pl}}$ and $\alpha_\nu = \alpha_{\rm{th}} +
\alpha_{\rm{pl}}$.  Rewriting the equation in terms of the source
function $S_\nu = j_\nu / \alpha_\nu$ and optical depth $\tau(s) = -
\int_s^{s_{out}} \alpha ds^\prime$, where $s_{out}$ is the point of
intersection of the ray with the outer boundary of the flow, equation
(\ref{eq:radtr2}) becomes
\begin{equation}
  \label{eq:radtr4}
\pm \frac{\partial{I_\nu}}{\partial{\tau}} = I_\nu - S_\nu, 
\end{equation}
where the combined source function is
\begin{equation}
  \label{eq:stot1}
  S_\nu = \frac{j_\nu}{\alpha_\nu} = \frac{j_{\rm{th}}+j_{\rm{pl}}}
   {\alpha_{\rm{th}}+\alpha_{\rm{pl}}} = \frac{S_{\rm{th}}}{1 + 
  \alpha_{\rm{pl}}/ \alpha_{\rm{th}}} + 
   \frac{S_{\rm{pl}}}{1 + \alpha_{\rm{th}}/ \alpha_{\rm{pl}}}. 
\end{equation}

In the numerical calculations reported below, we integrate equation
(\ref{eq:radtr4}) using the formal solution and the appropriate
boundary conditions for a non-illuminated atmosphere (see Mihalas
1978).  We carry out the integral along rays with impact parameters up
to $\sim 2000$ Schwarzschild radii, beyond which the temperature of
the electrons becomes too low for significant synchrotron emission.
Because of the very steep dependence of the synchrotron emissivity on
photon frequency, magnetic field strength, and electron temperature
and density, we use an adaptive step size for the radiative transfer
integral. The total flux is obtained by integrating over all impact
parameters. We validated the implementation of our numerical algorithm
by comparing its output to analytic solutions of the radiative
transfer equation in uniform media. The discrepancy between the
numerical and analytic solutions was $\lesssim 0.5 \%$ for the cases
considered.

In all model ADAF spectra published so far, the transport of radiation
has been calculated using an approximate method based on concentric
shells (Narayan, Barret, \& McClintock 1997a). In Figure 1, we
compare, for a typical flow, the exact spectrum obtained by formally
solving the above transfer equation to that obtained with the
previous approximate method. We see that there is a fairly good
agreement between the two methods, but with some differences. The most
prominent difference is a shifting of the peak toward higher frequencies
in the exact calculation, as well as some
broadening. This is probably due to the poorer resolution of the
concentric shell method which becomes a limiting factor close to the
black hole. There is also a slight offset at lower frequencies,
probably again due to poor resolution.  Note that there is a second
peak at high frequencies ($\gtrsim 10^{14}$ Hz) in the spectrum
calculated with the approximate method.  This is caused by the inverse
Compton scattering of soft photons, a process which is not included in
the radiative transfer code described here.  The remaining features of
the two spectra are quantitatively consistent.

\vbox{ \centerline{ \psfig{file=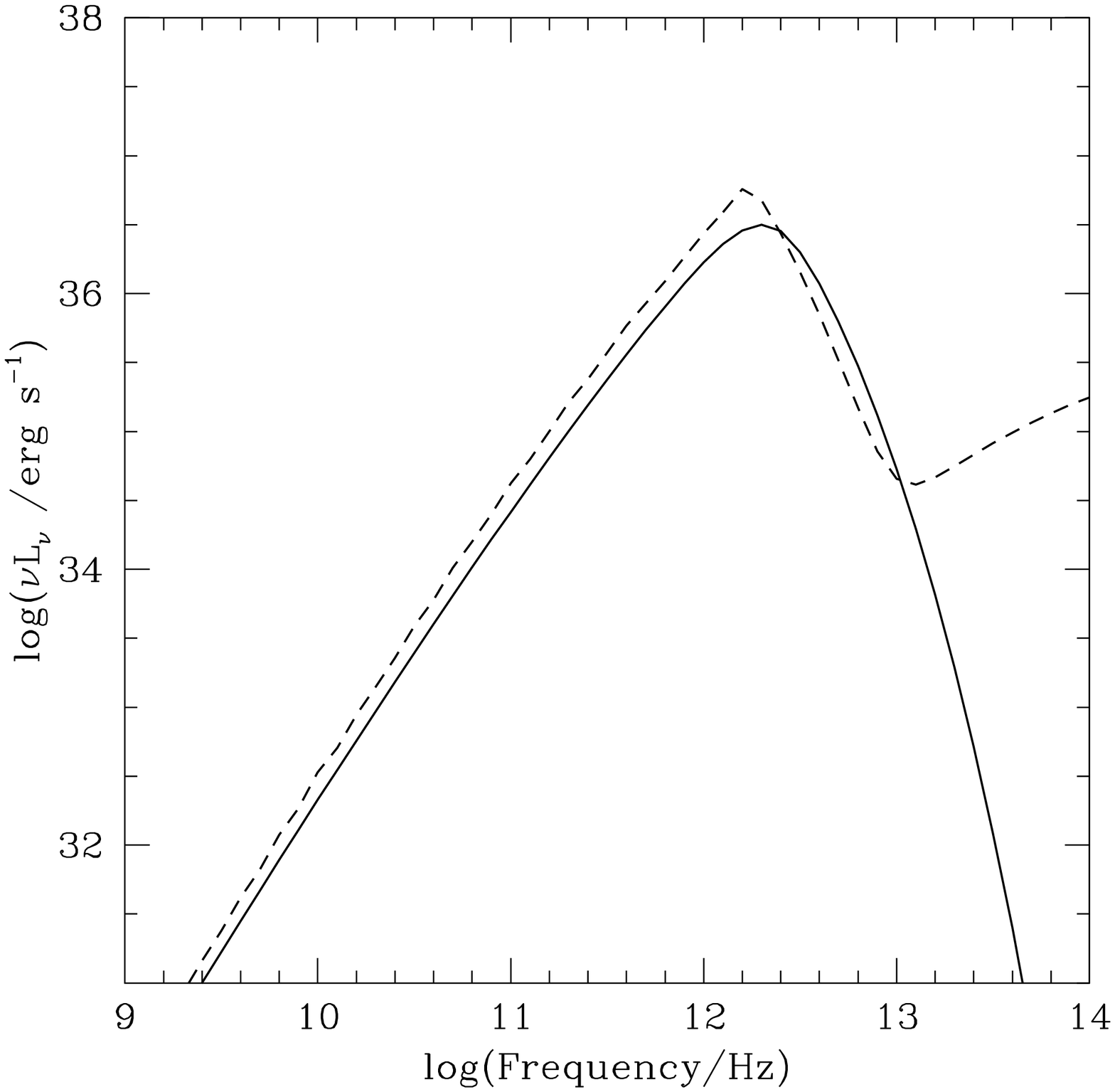,width=9.0truecm} }
\figcaption[]{ \footnotesize The solid line shows a typical spectrum
calculated with the radiative transfer method employed in this paper.
The model parameters are those used for Sgr~A$^*$: $m_6=2.5$,
$\dot{m}_{-3}=0.1$, and other ADAF parameters as specified in the
text.  The dashed line is an approximate spectrum of the same model,
calculated by a simplified method described in Narayan et al.
(1997a). The secondary rise at $\nu \gtrsim 10^{13}$ Hz is due to
Compton scattering, which is not included in the exact calculation.  }
}
\vspace*{0.5cm} 

\section{Numerical Results}

\subsection{Parameter Study}

In this section, we study the effects of an extended power-law
electron distribution on the synchrotron emission spectra of ADAFs.
Figure 2 shows the various components of the spectrum of a typical
hybrid model, with $p=3.5$, $\eta=1 \%$, $m_6=2.5$, and
$\dot{m}_{-3}=1$. Compared to the spectrum of a pure thermal model
(dashed line), we see two primary effects due to the power-law
electrons: (i) there is a prominent shoulder of optically thick
emission at low frequencies and (ii) there is an extended power-law
tail of optically thin emission at high frequencies. In between these
two features there is a region of the spectrum where the thermal peak
dominates. These features were first identified by Mahadevan (1998)
for a specific model. We find that they are universal for hybrid models. 

\vbox{ \centerline{ \psfig{file=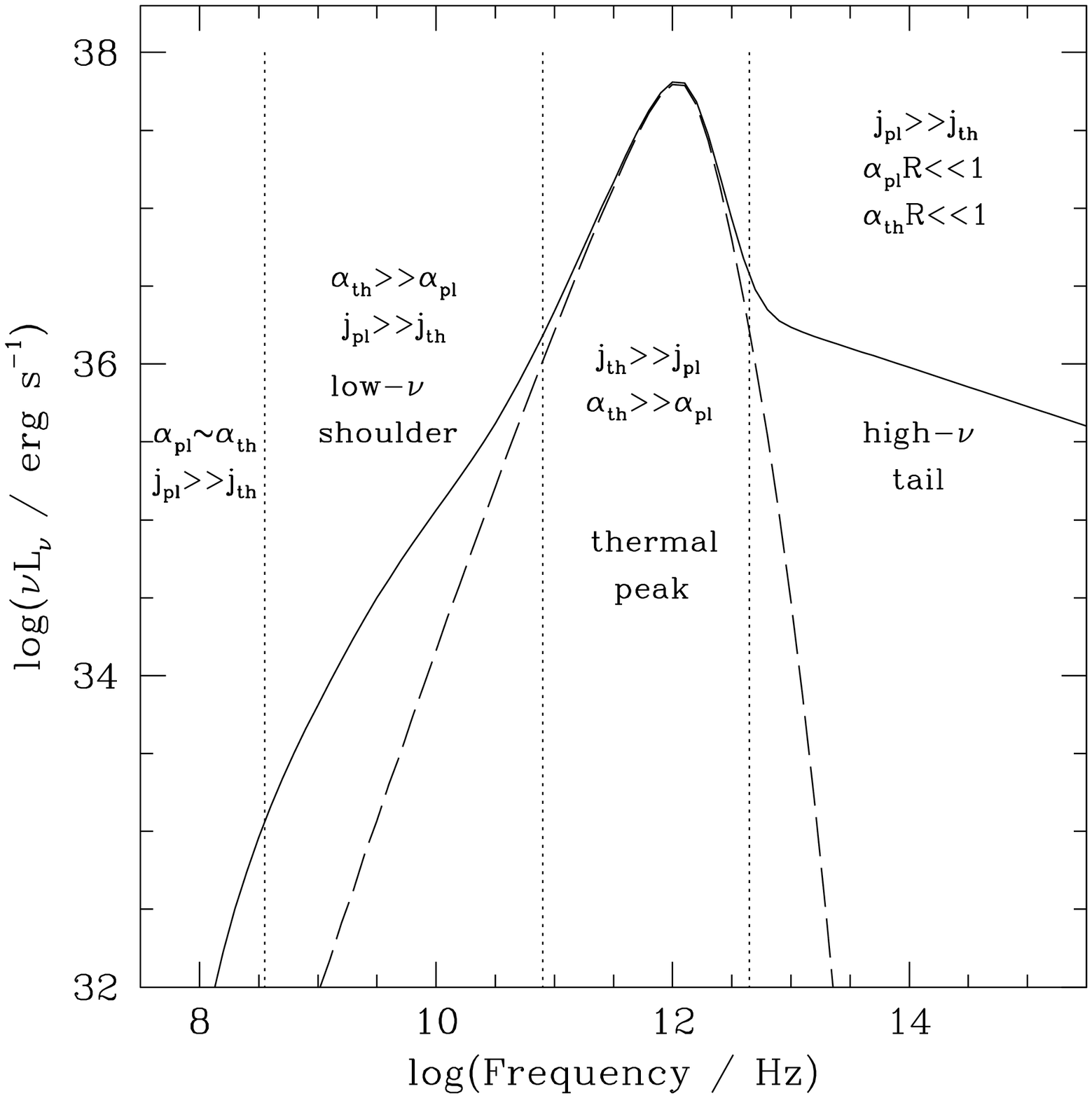,width=9.0truecm} }
\figcaption[]{ \footnotesize Regions of a hybrid synchrotron spectrum
  labeled according to the dominant emitting and absorbing electron
  population and the resulting spectral shape. The different segments
  and the transitions are present in all hybrid synchrotron spectra.
  }}
\vspace*{0.5cm} 

To understand the results, we note that the spectra of hybrid
populations are determined by a competition between $j_{\rm{pl}}$ and
$j_{\rm{th}}$ and between $\alpha_{\rm{pl}}$ and $\alpha_{\rm{th}}$,
each having different radial and frequency dependences. Since the
emission at different frequencies arises from different radii in the
flow, the relative importance of $\alpha_{\rm{pl}}$ to
$\alpha_{\rm{th}}$ and $j_{\rm{pl}}$ to $j_{\rm{th}}$ at that
particular frequency and radius determines the local behaviour of the
spectrum. In the less steep segment of the low-frequency shoulder, the
emission from the power-law population (which is more efficient at low
frequencies) exceeds that of the thermal electrons, while absorption
is still mostly dominated by the more numerous thermal electrons. As a
result, this segment of the spectrum assumes a shape roughly described
by the rather unusual source function $S =
j_{\rm{pl}}/\alpha_{\rm{th}}$. At still lower frequencies, where the
emission comes from larger radii in the flow, the contribution of the
non-thermal electrons to the absorption becomes non-negligible and the
spectrum falls more steeply with decreasing $\nu$.

In the region of the thermal peak, we have both $j_{\rm{th}} \gg
j_{\rm{pl}}$ and $\alpha_{\rm{th}} \gg \alpha_{\rm{pl}}$, with
$\alpha_{\rm{th}} R \gg 1$ for frequencies smaller than the peak
frequency and $\alpha_{\rm{th}} R \ll 1$ for frequencies higher than the
peak frequency. The spectrum is essentially the same as for a purely
thermal model (dashed line).  (But, note that for sufficiently high
$\eta$, the synchrotron spectrum assumes an entirely non-thermal 
character and this region too can be dominated by the power-law electrons,
with the peak luminosity becoming higher and the peak flattened).
Beyond the thermal peak, there is no self-absorption either by thermal
or power-law electrons $(\alpha_{\rm{th}} R, \alpha_{\rm{pl}} R \ll 1)$
and the emission is optically thin.  Here we find an extended
power-law tail dominated entirely by the non-thermal population. This
segment of the spectrum has the familiar form $\nu L_\nu \sim
\nu^{-(p-3)/2}$ which gives a rising spectrum with increasing
frequency for $p<3$ and a falling spectrum for $p>3$.

\vbox{ \centerline{ \psfig{file=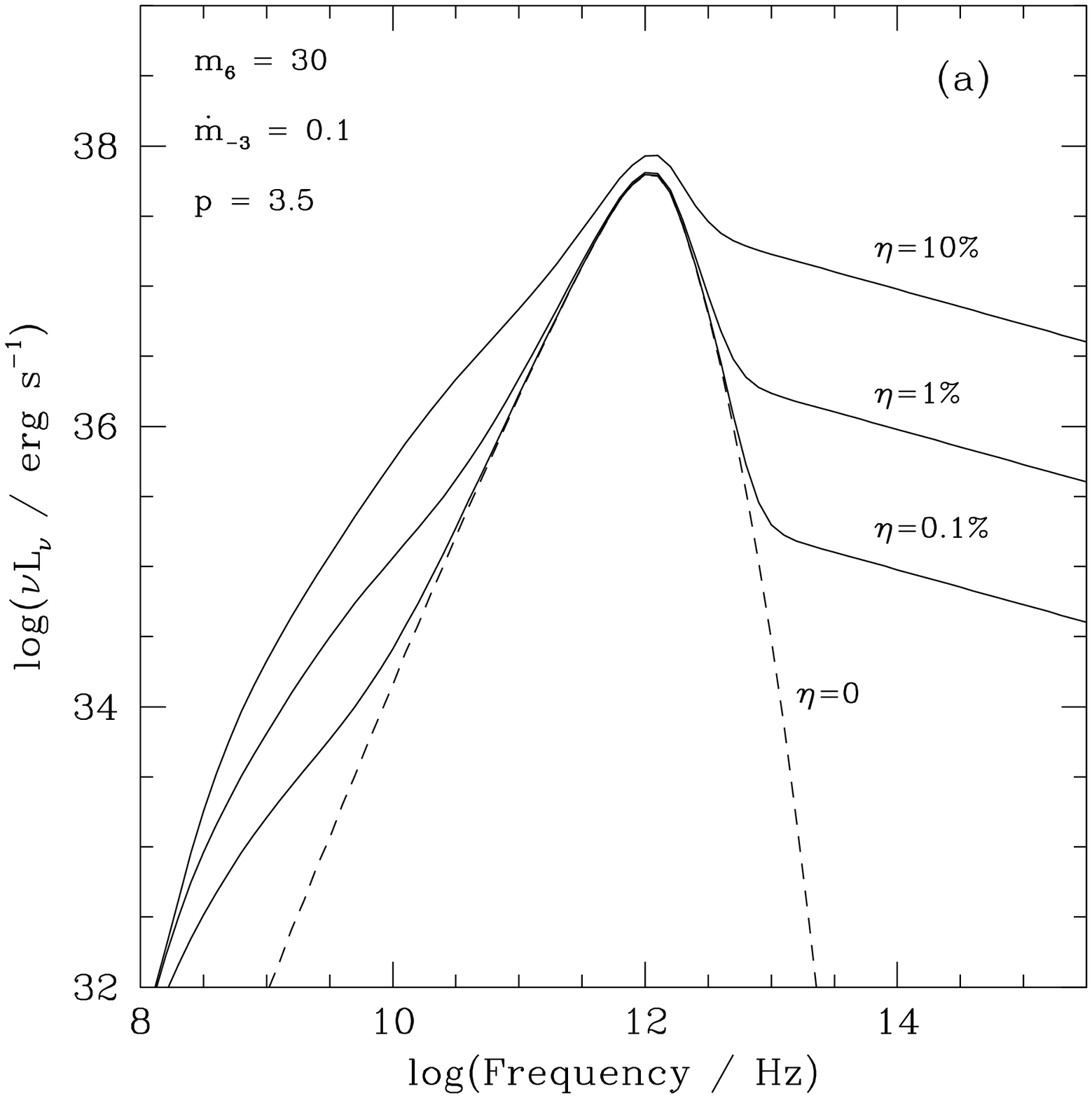,width=5.2truecm}
    \psfig{file=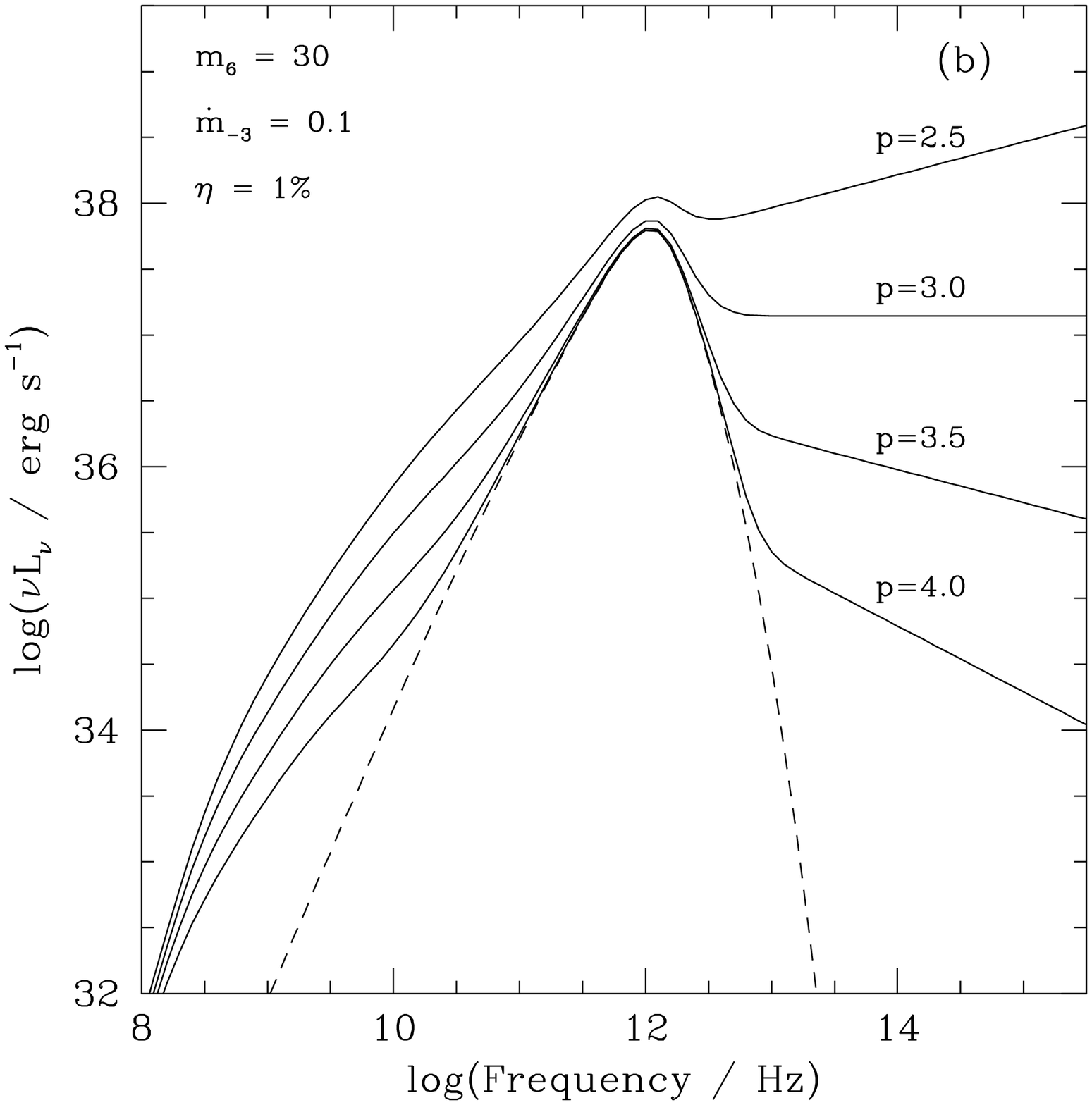,width=5.2truecm} } \centerline{
    \psfig{file=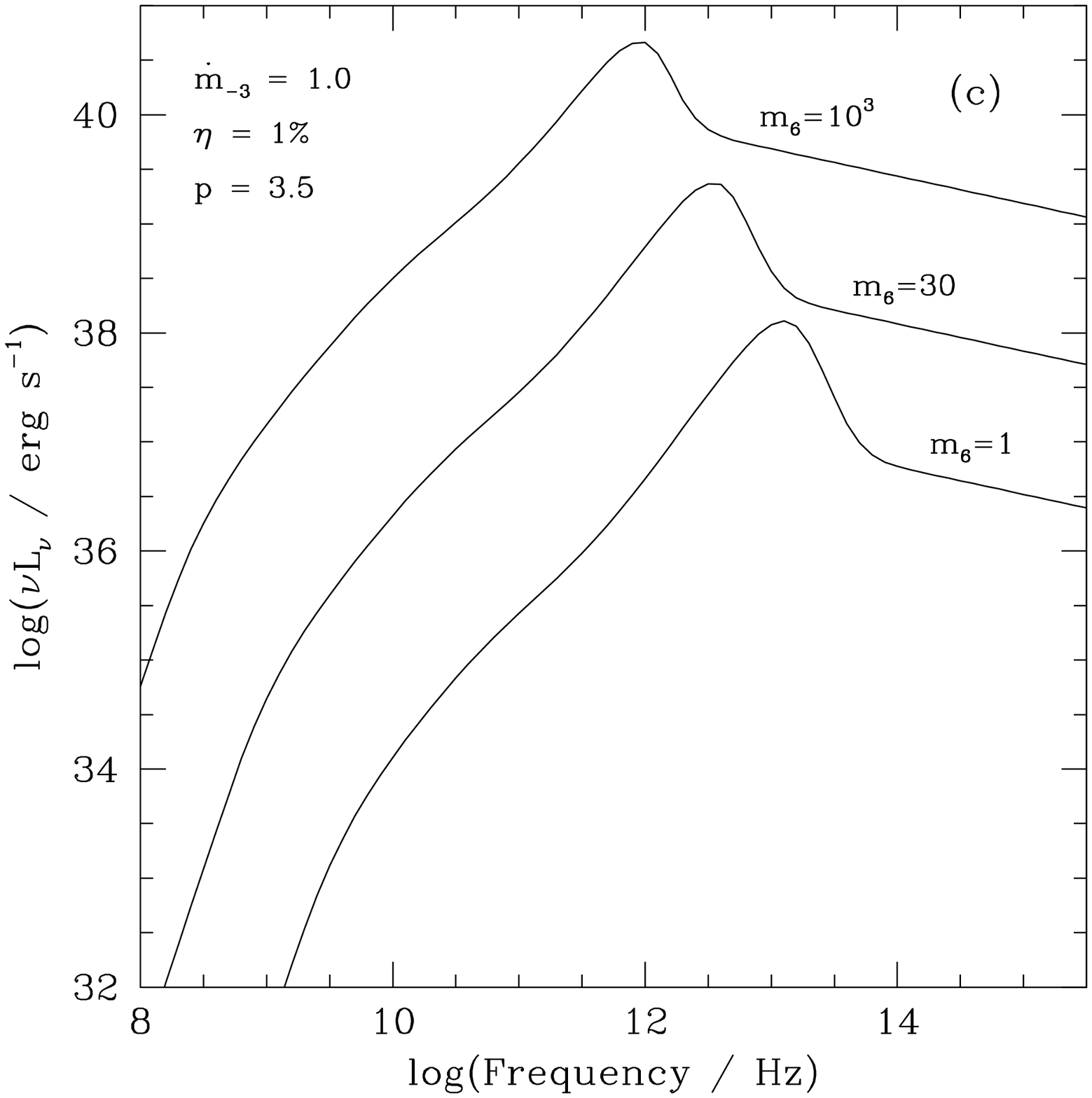,width=5.2truecm}
    \psfig{file=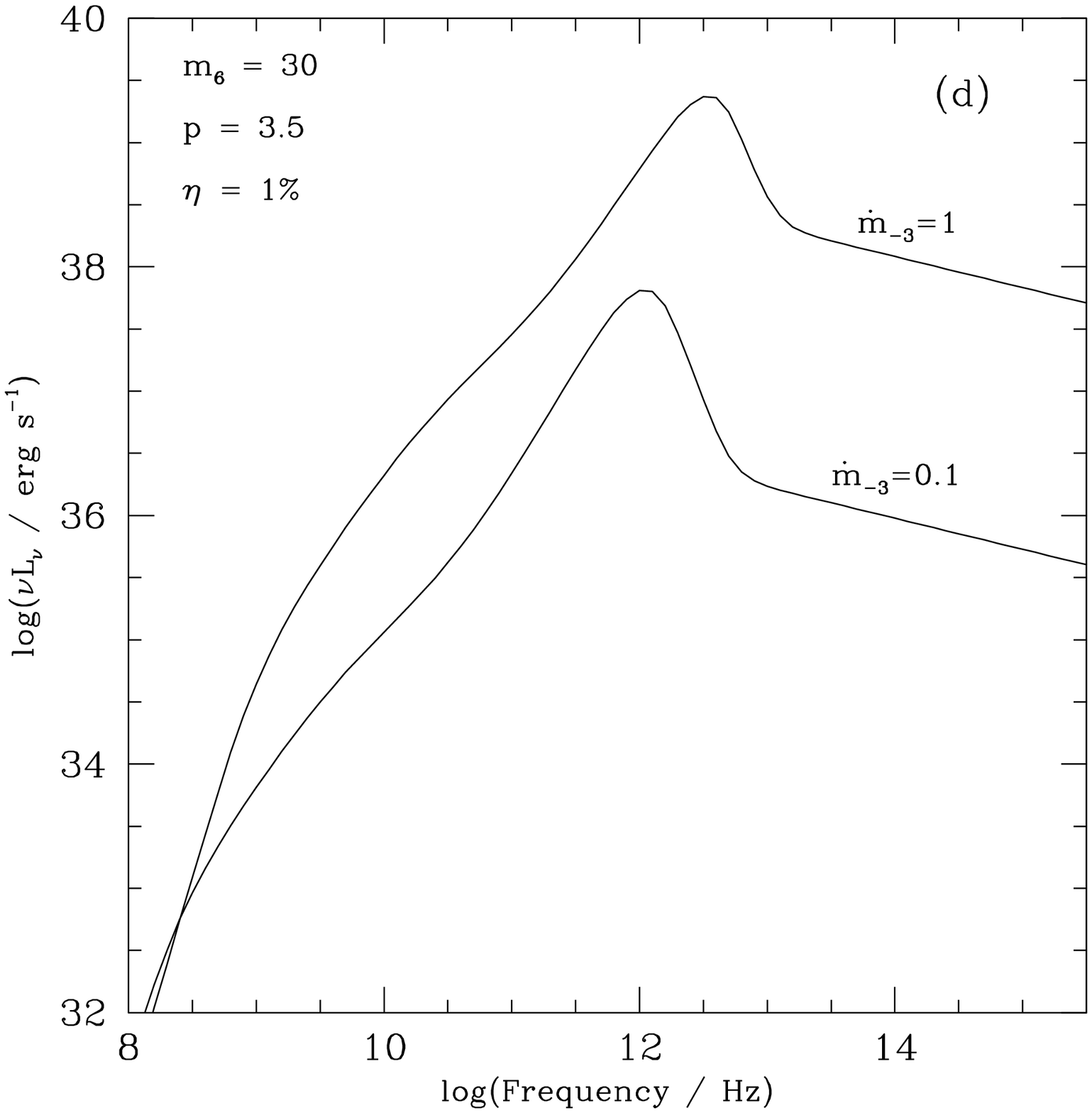,width=5.2truecm} }
  \figcaption[]{\footnotesize Variations in the synchrotron spectrum
    as a function of (a) the fraction of non-thermal energy content
    $\eta$, (b) the power-law index $p$, (c) the black hole mass in
    units of $10^6 M_{\odot}$, and (d) the mass-accretion rate in
    units of $10^{-3} \dot{M}_{\rm{Edd}}$.  }} 
\vspace*{0.5cm}

Figures 3a-d show how the spectrum of a hybrid model depends on the
various parameters. In Figure~3a, the energy content of the power-law
population $\eta$ is varied from $0 \%$ (purely thermal) to $10 \%$;
we see that the range and normalization of the low-frequency shoulder
increase linearly with increasing $\eta$. The normalization of the
high-frequency tail also depends linearly on the number of electrons in
the power-law distribution and thus varies in an obvious way with
$\eta$.

In Figure~3b, we fix $\eta$ at $1 \%$ and vary the power-law index
$p$. Note that the overall luminosity of both the low-frequency
shoulder and the high-frequency tail show considerable dependence on
$p$. This is not unexpected. When $p$ is small, there are more
particles both at intermediate Lorentz factors $(\log \gamma = 1-2)$,
which give rise to the low-frequency shoulder, as well as at large
$\gamma$, which emit the high-frequency tail (see $\S 4$).  Therefore,
small values of $p$ give more emission in both these segments of the
spectrum. The high-frequency tail has a spectral slope equal to
$-(p-3)/2$ and therefore has a strong dependence on $p$.

Two striking qualitative results in Figures 3a and 3b need to be
highlighted.  First, even very small values of the non-thermal energy
content $\eta$ give rise to significant excess super-thermal emission
at low frequencies. For instance, even if only $1 \%$ of the total
electron energy is in non-thermal electrons, there can be several
orders of magnitude higher luminosity at frequencies below the thermal
peak.  Even more striking is the fact that the spectral shape of the
low frequency shoulder is nearly independent of $\eta$ and $p$, when
both these parameters are independent of radius, while the
normalization depends on both (see Appendix A for an analytic
expression for the spectral slope of this segment). This leads to a
degeneracy between $\eta$ and $p$ in models of the low-frequency
emission, so that it is not possible to distinguish between different
combinations of $\eta$ and $p$, by studying spectral data below the thermal
peak only.  The degeneracy can be lifted with data on the
high-frequency tail whose slope has a strong dependence on $p$.

We finally consider the effect of changing the black hole mass and the
mass-accretion rate. We find that the relative excess emission due to
non-thermal electrons increases rather weakly with increasing black
hole mass (Figure 3c). This is because the thermal peak moves to lower
frequencies for higher $m_6$ (Mahadevan 1997) and at these frequencies
the effect of non-thermal electrons is more prominent.  The
non-thermal contribution to both the low and high frequency emission
increases only weakly with increasing $\dot{m}_{-3}$ (Figure 3d).

\subsection{Application to Sgr~A$^*$}

Over the last decade, many models have been developed to explain the
radio spectrum of Sgr~A$^*$.  This source is believed to be an
accreting supermassive black hole at the center of our Galaxy.  Nearly
all the published models invoke synchrotron radiation from
relativistic or quasi-relativistic electrons.  Melia (1992) considered
emission by thermal electrons in a spherical accretion flow and showed
that the resulting cyclo-synchrotron emission is consistent with the
broad features of the observed spectrum.  Narayan, Yi, \& Mahadevan
(1995) and Narayan et al. (1998a) developed ADAF models of Sgr~A$^*$
which included rotation, viscosity and a two-temperature plasma, and
obtained similar results, again with a purely thermal distribution of
electrons.  Beckert \& Duschl (1997) and Falcke \& Biermann (1999)
considered non-thermal models, while Mahadevan (1998) analyzed a
specific hybrid model in which the non-thermal electrons are produced
by pion decay.  This section is a generalization of Mahadevan's
work.

In Figure 4 we apply our hybrid emission model to Sgr~A$^*$. We take
$m_6 = 2.5$ (Eckart \& Genzel 1997; Ghez et al. 1998) and adjust
$\dot{m}_{-3}$ in order to fit approximately the thermal peak.  The
data shown are the same as in Narayan et al.  (1998a).  Figure 4 shows
four models with $p=2.5$, $\eta = 0.05\%$; $p=3.0$, $\eta = 0.2\%$;
$p=3.5$, $\eta = 0.5\%$; and $p=4.0$, $\eta =1\%$.  We first note that
the agreement with data at low frequencies is significantly better
with these hybrid models (dashed lines) than with a purely thermal
model (solid line), as was first shown by Mahadevan (1998).

The two major results pointed out in $\S 3.1$ are evident in this
figure.  First, there is no unique solution for the parameters $p$ and
$\eta$. The four hybrid models shown in Figure 4 give
indistinguishable spectra below the thermal peak, rendering it
impossible to determine $p$ and $\eta$ from low-frequency spectral
data alone. Second, $\eta$ is extremely small in all models. A very
small fraction of the energy in non-thermal electrons, with $\eta$ at
most $1 \%$, is sufficient to produce all the observed emission at low
frequencies.  This means that the non-thermal electrons are a minor
perturbation on the electron population, which is itself a minor
perturbation on the more dominant ion population.  We are therefore
consistent when we compute the gas dynamics with a purely thermal
model and ignore the non-thermal electrons for the dynamics.  Further
justification of this assumption as well as implications of this
constraint are discussed in $\S 5$.

We can obtain additional constraints on the parameters $\eta$ and $p$
in Sgr~A$^*$ by studying the infrared data. Sgr~A$^*$ is quiet at
infrared wavelengths, with a current upper limit of $10^{35}$ erg
$\rm{s}^{-1}$ on the luminosity at 2.2~$\mu$m (Eckart \& Genzel 1997;
Ghez et al. 1998). Since $\nu L_{\nu}$ decreases with increasing
frequency for $p>3$, we cannot use IR data to constrain $\eta$ very
strongly in these cases. On the other hand, if $p<3$, electrons in an
extended power-law distribution produce significant emission in the
infrared.  For example, if $p=2.5$, we find that the maximum allowed
fraction of energy in non-thermal electrons is $\eta=0.05 \%$. Tighter
bounds on the infrared flux will constrain the parameters $\eta$ and
$p$ even more strongly.

We note that imposing a maximum Lorentz factor $\gamma_{\rm{max}}$ on
the power-law electron distribution also has the effect of suppressing
the high-frequency emission. Therefore, for $p < 3$, we could
alternatively use the IR data to constrain $\gamma_{\rm{max}}$ rather
than $\eta$. The $2.2~\mu$m emission is produced predominantly by
electrons with $\log \gamma \gtrsim 3$ (see $\S 4$) placing a maximum
Lorentz factor at $\gamma_{\rm{max}} \sim 10^3$, if $p < 3$ and $\eta
> 0.05 \%$.

Finally, we have considered ADAF models with strong outflows following
the ideas described in Blandford \& Begelman (1999), Di Matteo et al.
(1999), and Quataert \& Narayan (1999). We find that the constraints
on $\eta$ and $\gamma_{\rm{max}}$ obtained above do not depend
strongly on the presence of winds.

\vbox{ \centerline{ \psfig{file=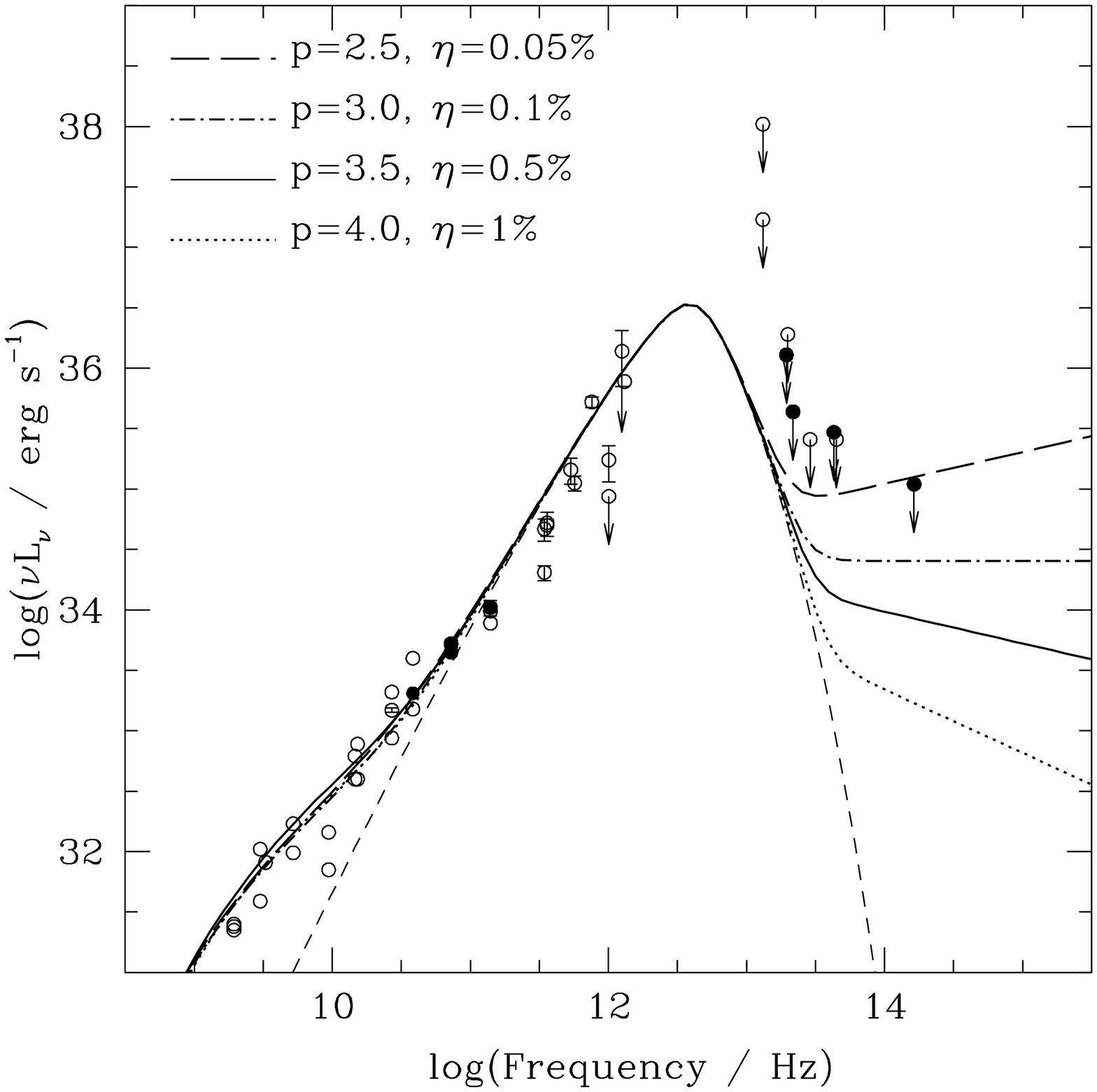,width=9.0truecm} }
\figcaption[]{ \footnotesize Comparison of spectral models for
  Sgr~A$^*$ with radio and IR data. The dashed curve shows the
  spectrum when the electrons are purely thermal.  The other four
  curves show spectra from hybrid populations with the following
  combinations of parameters: $p=2.5$, $\eta = 0.05\%$; $p=3.0$, $\eta
  = 0.2\%$; $p=3.5$, $\eta = 0.5\%$, and $p=4.0$, $\eta =1\%$.  }}
\vspace*{0.5cm} 

\subsection{Image sizes and shapes}

We now investigate the effect of a power-law electron population on
the size and shape of the radio image of an ADAF. As a typical
example, we consider an ADAF model of Sgr~A$^*$ with $m_6=2.5$ and $
\dot{m}_{-3} = 0.1$ and take a hybrid electron distribution with $p =
3.5$, $\eta = 0.5\%$ which agrees well with the observed spectrum as
shown in Figure~4. Preliminary size measurements of Sgr~A$^*$ are
available at two frequencies (Lo et al. 1998, Krichbaum et al. 1998)
and shape measurements may be possible in the near future.  We note
that although we use the parameters for a specific source, the
qualitative results apply equally well to other massive black holes
with different accretion rates.

We define the image size as twice the radial distance from the center
of the image to the point at which the specific intensity falls to
half the value of the central intensity, i.e., the FWHM of the radio
map. Figure~5 shows the predicted FWHM of the image as a function of
frequency. The dashed curve corresponds to the case when all the
electrons are thermal and the solid line to the case of a hybrid
energy distribution. The two data points for Sgr~A$^*$ are from Lo et
al.  (1998) at 7~mm and Krichbaum et al. (1998) at 1.4~mm. Note that
our radiative transfer code is one dimensional and can only handle
spherical models, though ADAFs in general are oblate and are likely to
appear elliptical in projection (Narayan \& Yi 1995a). In the
comparison with observations, we use only the measured long axes of
the images.

\vbox{ \centerline{ \psfig{file=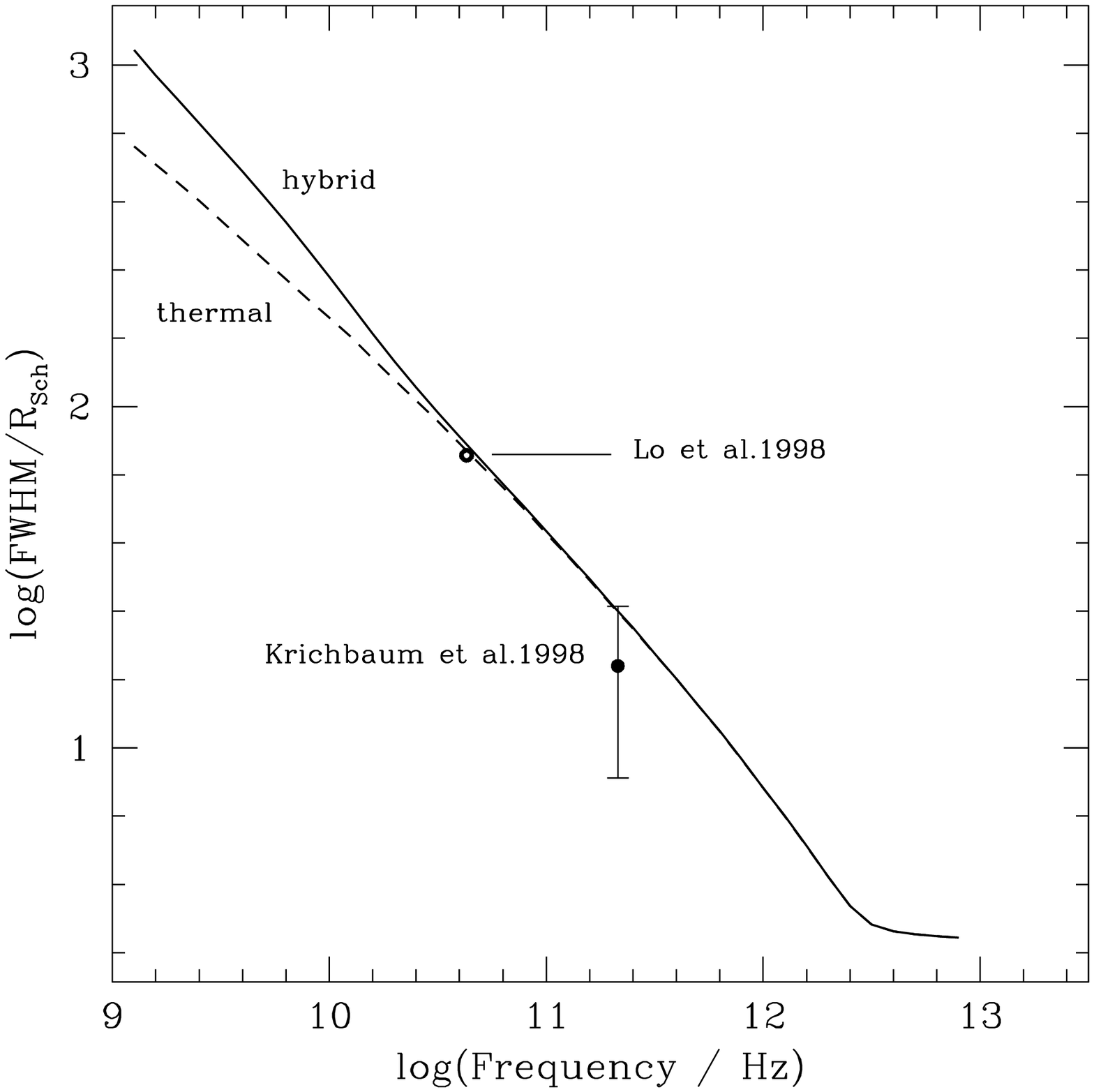,width=9.0truecm} }
\figcaption[]{ \footnotesize The solid line shows the FWHM image size
  corresponding to one of the models described in Figure 4, with
  $p=3.5$ and $\eta=0.5\%$.  This is compared with two size
  measurements of the Galactic Center source Sgr~A$^*$ at 1.4mm
  ($\nu=214$ GHz) and 7mm ($\nu=43$GHz). We do not show error bars of
  the 7mm measurement (Lo et al. 1998) because of the oblateness of
  the observed image which we have not included in our models.  The
  dashed line corresponds to a pure thermal model. }}
\vspace*{0.5cm}

The effect of power-law electrons is to enlarge the image size at long
wavelengths $(\nu < 10^{11} \rm{Hz})$, where the excess non-thermal
emission, which extends outward of the surface of unit optical depth,
is most prominent. This is a general result which holds true for all
model parameters we have studied. The other related result is the
steepening of the dependence of the image size on the frequency at
long wavelengths. While for thermal electrons we find
\begin{equation}
\rm{FWHM/R_{Sch}} \sim \lambda^{0.7},
\end{equation}
when we include power law electrons, we find
\begin{equation}
\rm{FWHM/R_{Sch}} \sim \lambda^{0.9}.
\end{equation}
This is in agreement with the preliminary result obtained by Lo et al.
(1998) when they combined their measurement of the intrinsic size at
7~mm with the results of Krichbaum et al. (1998) at 1.4~mm, although
the thermal and hybrid slopes are not distinguishable with the current
data.

We also calculate the effective brightness temperature $T_b$ at each
frequency as $T_b = L_{\nu} c^2 / (8 \pi^2 k_B \nu^2 \rm{FWHM}^2)$.
Figure~6 shows the result. For the thermal model, $T_b$ measures the
electron temperature at the photosphere. Since the photosphere moves
out with decreasing frequency, the brightness temperature falls. For
the hybrid model, at lower frequencies, the contribution of the
non-thermal electrons is large, the spectrum is non-thermal, and the
resulting brightness temperature increases.  This is another clear
signature of a synchrotron-emitting hybrid electron population.

\vbox{ \centerline{ \psfig{file=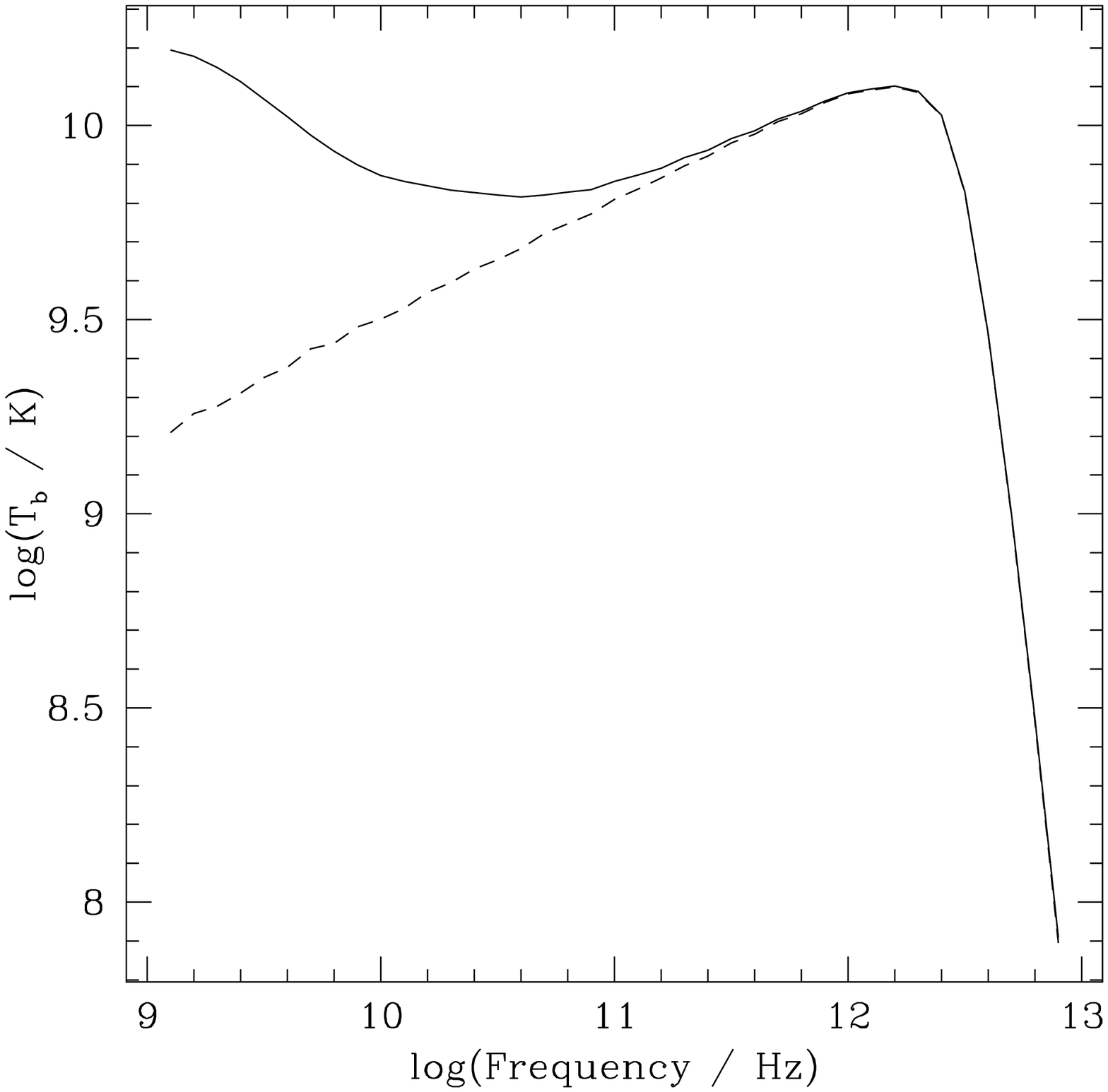,width=9.0truecm} }
\figcaption[]{ \footnotesize The predicted brightness temperature of
  Sgr~A$^*$ as a function of frequency. The flux was taken from
  Figure~4 and the size estimate from Figure~5. The dashed line
  corresponds to the purely thermal model and the solid line to a
  hybrid model with $p=3.5$ and $\eta=0.5\%$. }} 
\vspace*{0.5cm}

We finally study the effect of the non-thermal electrons on the shape
of the radio image. Figure~7 shows the variation of the brightness
temperature $T_b$ across the source, for the thermal and hybrid models
of Sgr~A$^*$ described above.  We consider two wavelengths, 3.6~cm and
7~mm. The most striking feature of the hybrid case is the limb
brightening seen at long wavelengths, a feature which is absent in the
thermal models. Thus, if the accretion flow contains non-thermal
electrons, its image would look like a shell rather than a disk. This
is again due to the different radial dependences of absorption (which
is predominantly thermal) and emission (which is mostly non-thermal)
at the frequencies where the non-thermal shoulder appears in the
energy spectrum.  Because the total absorption falls off more steeply
away from the center than the total emissivity, the image appears
brighter for a range of impact parameters away from the center than at
the center. In addition, due to the overall increase in the intensity
of the emerging radiation in this same frequency range, the image
looks brighter overall, with the brightness temperature $T_b$ of the
hybrid case increasing to twice the thermal value in the center of the
image at 3.6 cm.  The limb brightening, along with the enhanced
overall brightness at long wavelengths are signatures of a non-thermal
population which may be accessible to observations. Unfortunately, in
the case of Sgr~A$^*$, precisely at the wavelengths where the effects
are strongest, interstellar scattering blurs the observed image (e.g.,
Lo et al. 1998). It may be worthwhile to look for these effects in
other sources where the scattering is less severe.

\vbox{ \centerline{ \psfig{file=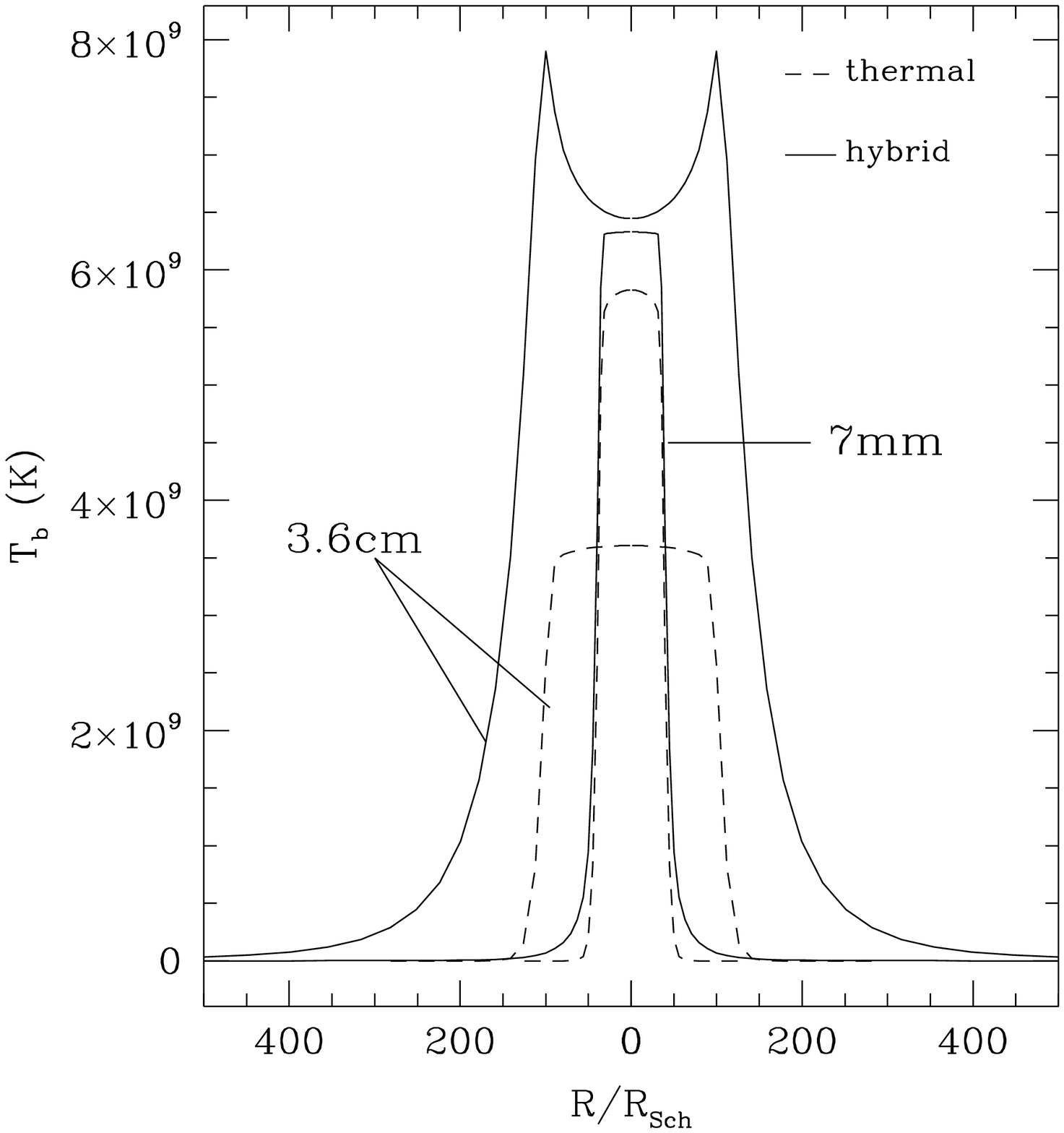,width=9.0truecm} }
\figcaption[]{ \footnotesize The predicted image shape of Sgr~A$^*$ at
  7mm and 3.6cm. The dashed line corresponds to the purely thermal
  model and the solid line to the hybrid model with $\eta=0.5\%$ and
  p=3.5. Note that the hybrid model has a higher brightness
  temperature at both wavelengths, and is significantly
  limb-brightened at 3.6cm.  }}
\vspace*{0.5cm}

\section{Electron Energy Distributions and the Shape of Synchrotron Spectra}

The numerical studies presented in $\S 3$ show that the presence of a
small population of power-law electrons causes universal modifications
to the spectrum: it introduces a shoulder at low frequencies and a
power-law tail at high frequencies. We now attempt to associate each
of these features of the hybrid spectrum with a specific range of
Lorentz factors of the non-thermal electrons. In doing so, we address
three important issues.

The first is the question of degeneracy: why is the low frequency
shoulder in the spectrum degenerate to a combination of the power-law
index $p$ and the energy content $\eta$ of the non-thermal electrons?
Understanding the source of non-uniqueness is especially important in
trying to extract the underlying electron distributions from spectral
data and determining what we can conclude $\it{uniquely}$ about these
distributions.

Second, since introducing even a very small fraction of power-law
electrons results in significant enhancement of the luminosity at low
frequencies, existing radio data can be used to constrain the
non-thermal energy fraction in these accretion flows. If only
a part of the electron energy distribution is responsible for the
excess emission, data can further constrain these specific parts of
the particle distribution.

The third question we address here is related to the absence of
significant IR emission from the accretion flow around Sgr~A$^*$
(Eckart \& Genzel 1997), in the frequency range where the extended
optically-thin non-thermal emission is expected.  Absence of emission
at these wavelengths provides information about the underlying
electron distributions by constraining either the power-law index $p$
or the maximum Lorentz factor $\gamma_{max}$ electrons can attain in
these accretion flows. Determining $\gamma_{max}$ in a particle
distribution may provide a better understanding of particle
acceleration and cooling processes and timescales in hot accretion
flows.

\subsection{Correspondence between electron distributions and photon spectra}

We first investigate the relationship between the Lorentz factor
$\gamma$ of an electron and the dominant frequency of the synchrotron
emission it produces in the presence of a dominant thermal electron
population.  At very low frequencies, synchrotron emission comes from
large radii in the flow where the temperatures are low, the
non-thermal emission and absorption are efficient, and
$\alpha_{\rm{pl}} \simeq \alpha_{\rm{th}}$. However, at higher
frequencies near and above the low-frequency shoulder in the spectrum,
$\alpha_{\rm{th}} \gg \alpha_{\rm{pl}}$, as discussed earlier. For
these frequencies, we can thus neglect $\alpha_{\rm{pl}}$. We make
this key simplification only in this section. It simplifies the
radiative transfer equation and allows us to compute separately the
contribution to the intensity from each value of $\gamma$ in the
non-thermal distribution.

We proceed by writing all quantities as integrals over the power-law
electron distribution,
\begin{equation}
j_{\rm{pl}} = \int n_{e}^{\rm{pl}}(r,\gamma) j(\gamma) d\gamma,
\end{equation}
\begin{equation}
I(\nu) \equiv \int n_e^{\rm{pl}}(\gamma) I_{\gamma}(\gamma) d\gamma, 
\end{equation}
and
\begin{equation}
j_{\rm{th}} = \frac{\int j_{\rm{th}} n_e^{\rm{pl}}(\gamma) d\gamma}
{\int n_e^{\rm{pl}}(\gamma) d\gamma},
\end{equation}
where $n_{e}^{\rm{pl}}(r,\gamma)$ is assumed to be a separable
function of $r$ and $\gamma$ given by
\begin{equation}
n_{e}^{\rm{pl}}(r,\gamma) = n_{r}^{\rm{pl}}(r) n_{\gamma}^{\rm{pl}}(\gamma) 
= N_{\rm{pl}}(r) (p-1) \gamma^{-p}.
\end{equation}
(Note that, for this analysis, the non-thermal distribution does not
have to be a power-law, though we have assumed this in order to
compare this analysis directly to our numerical results). 

\vbox{ \centerline{ \psfig{file=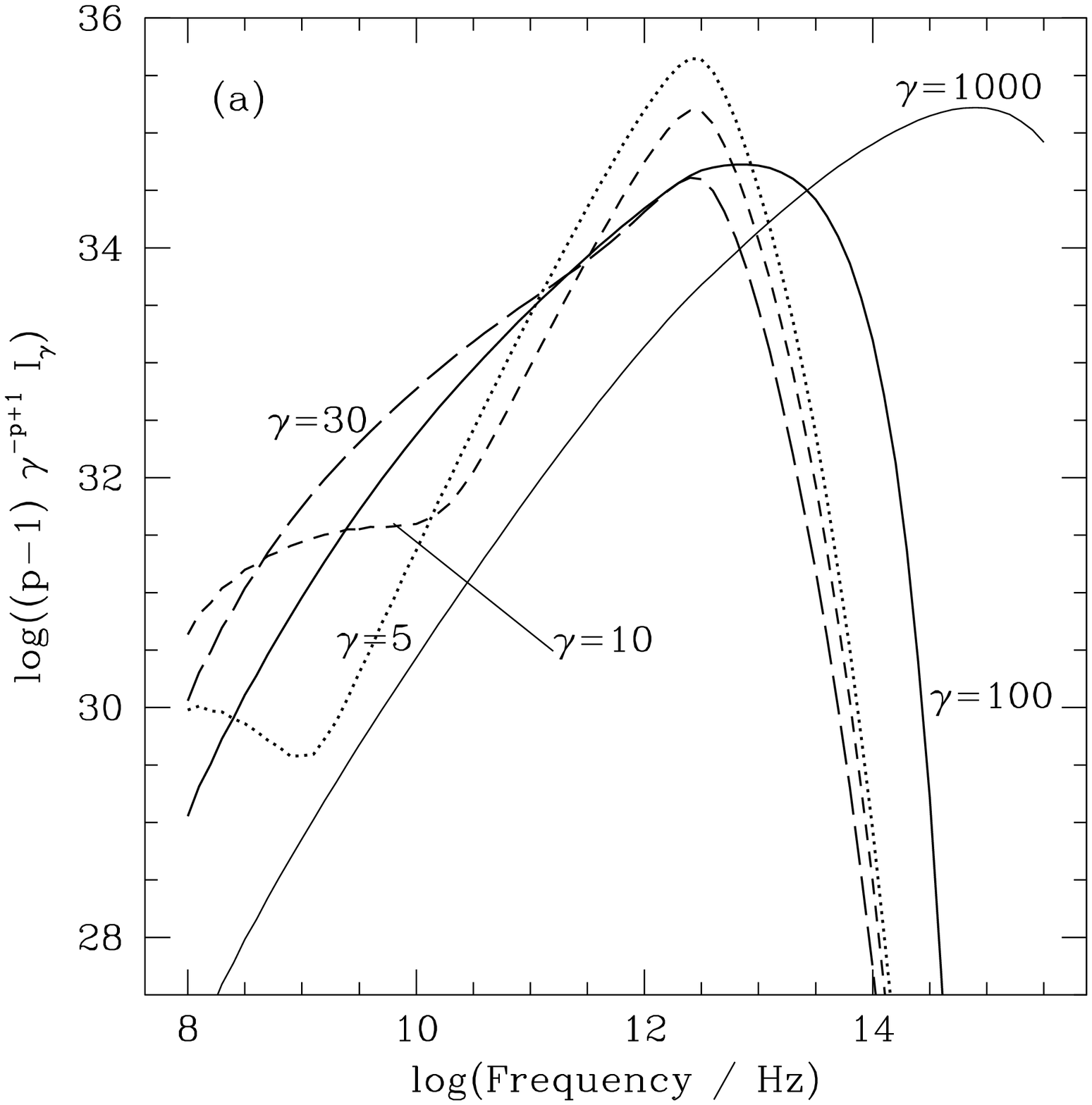,width=7.2truecm} }
  \centerline{ \psfig{file=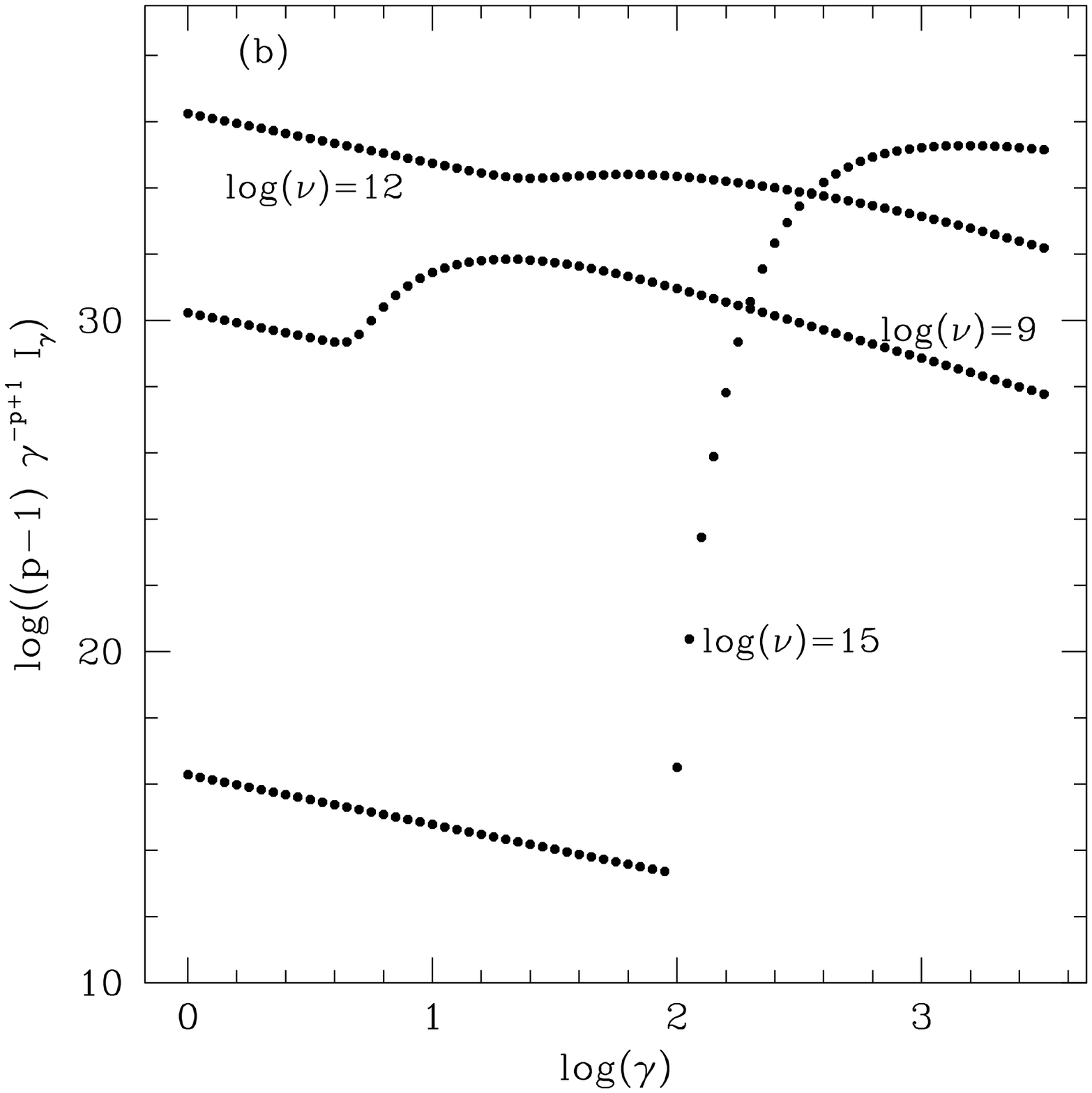,width=7.2truecm} } \centerline{
    \psfig{file=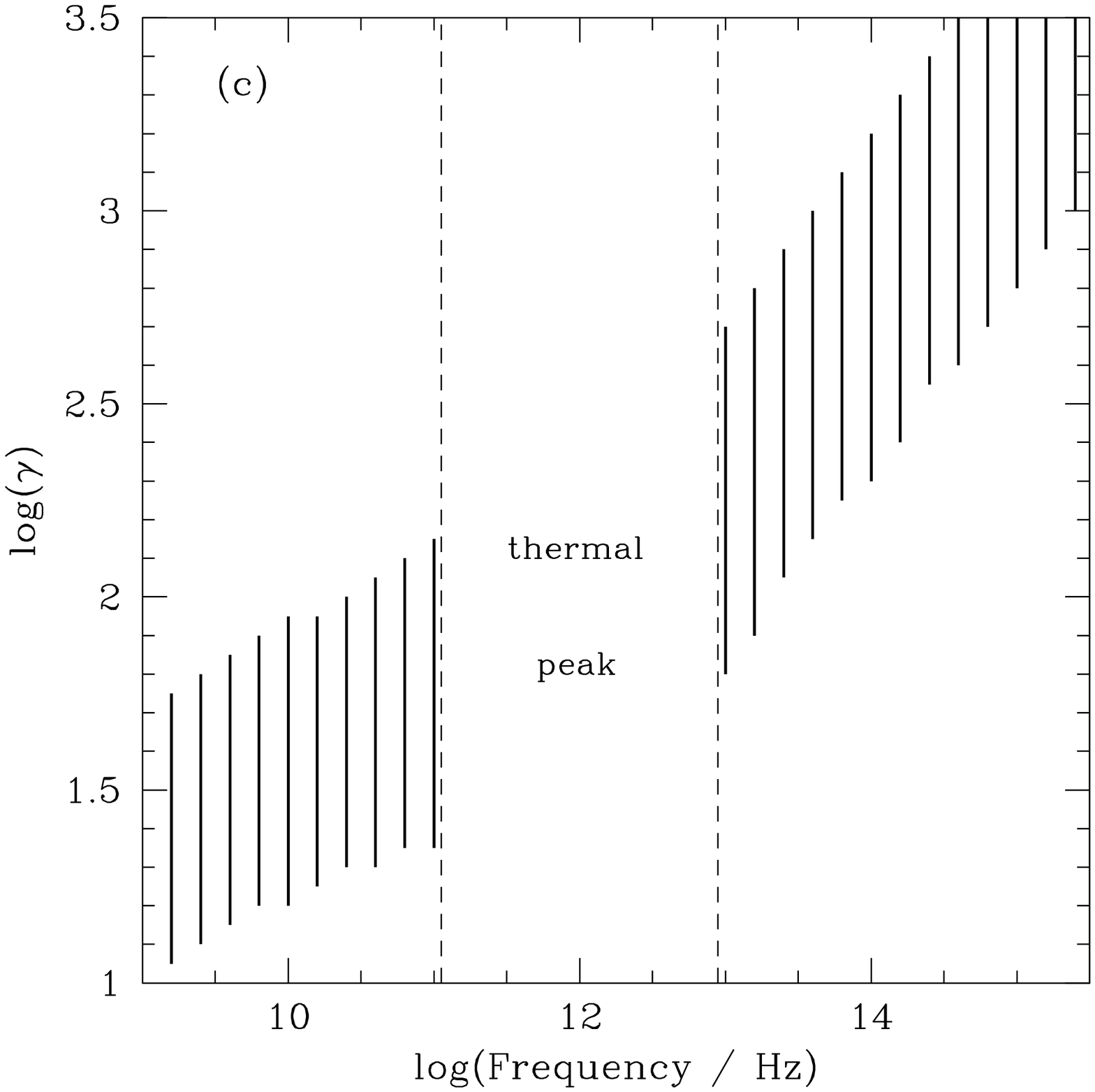,width=7.2truecm} } \figcaption[]{
    \footnotesize (a) The contributions to the hybrid spectrum of
    individual electron Lorentz factors: $\gamma=5, 10, 30, 100$, and
    $1000$. (b) The range of Lorentz factors contributing to the
    emission at $10^9$ Hz (low-frequency shoulder), $10^{12}$ Hz
    (thermal peak) and $10^{15}$ Hz (high-frequency tail).  (c) A
    further quantification of (b) showing the minimum and maximum
    Lorentz factors at which the integrand falls to half its maximum
    value at each frequency.  }}
\vspace*{0.5cm}

\newpage
\noindent Substituting equations (29)-(32) into the radiative transfer equation
and rearranging terms, we obtain
\begin{equation}
\int [\mu\frac{\partial I_{\gamma}(\gamma)}
{\partial x}+\alpha_{\rm{th}}(r) I_{\gamma}(\gamma) - 
n^{\rm{pl}}_r(r) j_{\gamma}(\gamma)- j_{\rm{th}}] 
 ~n^{\rm{pl}}_\gamma(\gamma) d\gamma = 0.
\end{equation}
For this integral to vanish for any non-thermal electron distribution,
the integrand must be identically zero, thus giving
\begin{equation}
\mu \frac{\partial I_{\gamma}(\gamma)}{\partial x} = 
-\alpha_{\rm{th}}(r) I_{\gamma}(\gamma) + n^{\rm{pl}}_r(r) 
j_{\gamma}(\gamma) + j_{\rm{th}}.
\end{equation}
We solve this equation for a wide range of Lorentz factors, $\log
\gamma = 0.1-3.6$, by choosing a specific radial profile of the
non-thermal electron density $N_{\rm{pl}}(r)$ corresponding to $p=3.5,
\eta=0.5 \%$. We study the contribution of each Lorentz factor to the
different parts of the spectra.  The total intensity for the power-law
distribution takes the form
\begin{equation}
I(\nu) \equiv (p-1)\int \gamma^{-p+1} I_{\gamma}(\gamma) d(\log\gamma),
\end{equation}
and therefore, in order to study the true contribution of each
$\gamma$ to the total spectrum, we plot the integrand $\gamma^{-p+1}
I_{\gamma}(\gamma)$ as a function of frequency for each Lorentz
factor. The result for $\gamma = 5, 10, 30, 100$, and 1000 (for
$p=3.5$) are shown in Figure~8a.  The curves give a good estimate of
the individual contribution of each Lorentz factor to the total
spectrum. Most of the contribution to the low-frequency shoulder comes
from electrons with $\gamma \sim 30-50$. Emission from electrons with
$\gamma \sim 100$ is already lower by an order of magnitude at those
frequencies, and the emission completely dies off beyond $\gamma=100$.
Contribution to the high-frequency tail, on the other hand, starts
around $\gamma \approx 100$ and increases with increasing electron
energy.

To get a more quantitative idea of which range of Lorentz factors
contributes to the three distinct regions of the spectrum, we show in
Figure~8b the above integrand as a function of $\gamma$ for 3
frequencies, namely $10^9$ Hz (in the low-frequency shoulder),
$10^{12}$ Hz (in the thermal peak), and $10^{15}$ Hz (in the
high-frequency tail. We see that the emission at $10^9$ Hz is
primarily from electrons with $\gamma \sim 10^{1.5}$; the emission at
$10^{12}$ Hz is mostly from $\gamma \sim 1$ (thermal electrons, while
the emission at $10^{15}$ Hz comes mostly from $\gamma \gtrsim 10^2$.

To quantify this effect further, we plot in Figure~8c the minimum and
maximum Lorentz factors for which the value of the integrand drops to
half its maximum value for each frequency. This gives us exactly the
contributing range of Lorentz factors to the emission at each
frequency in the low-frequency shoulder and the high-frequency tail.

We may summarize the results as follows. The low-frequency shoulder is
caused by a narrow range of electron Lorentz factors, $\log \gamma
\sim 1-2.$ The narrowness of the range explains why the spectra are
degenerate to different combinations of $p$ and $\eta$: a power-law
distribution with small values of $p$ and small $\eta$ has nearly the
same number of electrons in this narrow range of Lorentz factors as
one with a larger $p$ and a larger $\eta$, thus producing the same
emission in the shoulder. The high-frequency tail, however, is
produced by electrons with a wide range of Lorentz factors $(\log
\gamma \gtrsim 2)$ with higher frequencies coming from higher
$\gamma$. Intermediate Lorentz factors $\log \gamma \sim 2$ emit
predominantly at frequencies around the thermal peak and are
overpowered by the thermal emission, and therefore do not have an
observable feature in the hybrid spectrum.

\section{Energetics of the thermal and non-thermal populations}

In this section, we study the energy flow through the thermal and
non-thermal electron populations as well as the energy exchange
between these two populations through synchrotron self-absorption and
Coulomb collisions. Starting with the energy equations for the two
populations and a parametrization of the energy input into each, we
calculate the steady-state energy content of both as a function of
radius assuming that the shapes of the distributions are known a
priori. Note that if the acceleration mechanism is known, the particle
distributions can be calculated exactly (see, e.g., Nayakshin \& Melia
1998).  Here we simply assess the feasibility of a steady-state
thermal/power-law hybrid distribution.  We then extend our discussion
of hybrid synchrotron spectra to cases where the parameter $\eta$ is
not constant but is allowed to vary with radius.

\subsection{Energy Equations}

We start by writing the energy equations for the two populations.
Neglecting advection and diffusion which we estimate to be minor
corrections, the energy balance for the power-law population reads
\begin{equation}
  \label{eq:ennt}
  \frac{\partial E_{\rm{nt}}}{\partial t} = 0 = \delta_{\rm{nt}} 
\dot{E}_{\rm{visc}}(r) - \int j_{\rm{nt}}(\eta,p,r) d\nu d\Omega 
- \frac{E_{\rm{nt}}}{t_{ee}},
\end{equation}
while the thermal population obeys
\begin{equation}
  \label{eq:enth}
\frac{\partial E_{\rm{th}}}{\partial t} = 0 = \delta \dot{E}_{\rm{visc}}(r)
- \int \alpha_{\rm{th}}(B_{\nu}-I_{\nu}) d\nu d\Omega 
+ \frac{E_{\rm{nt}}}{t_{ee}},
\end{equation}
where $E_{\rm{th}}$ and $E_{\rm{nt}}$ are the steady state energy
contents of the thermal and power-law populations respectively,
$\delta$ is the fraction of viscous energy that heats the thermal
electrons, and $\delta_{\rm{nt}}$ is the corresponding fraction
injected into the power-law population ($\delta_{\rm{nt}} =
\eta_{\rm{inj}} \delta$, where $\eta_{\rm{inj}}$ is defined in $\S
2.2$); the time derivatives are set to zero in steady-state. The
energy exchange (thermalization) timescale due to Coulomb collisions
of high energy electrons with the thermal bath is denoted by $t_{ee}$.
The radiation term in equation~(\ref{eq:ennt}) corresponds to the
energy loss by the non-thermal electrons via optically thin
synchrotron emission; because the non-thermal absorption coefficient
is negligible throughout the flow, the non-thermal electrons do not
gain energy by absorption of synchrotron radiation. The corresponding
term in equation~(\ref{eq:enth}), on the other hand, describes the
heating of the thermal electrons by the local radiation field that is
in excess of the blackbody limit. When the radiation energy density is
a blackbody, there is locally no heating or cooling, but as the
radiation energy density is above this limit due to emission by the
non-thermal electrons, the thermal electrons are heated by absorbing
this total emission. Note that this is the same energy exchange
mechanism invoked in the synchrotron boiler process (Ghisellini,
Guilbert, \& Svensson 1988; Ghisellini, Haardt, \& Fabian 1993;
Ghisellini, Haardt, \& Svensson 1998).  The transport of energy due to
non-local synchrotron self-absorption is likely to be small and have
little effect on the spectra as the photon mean free path becomes very
short very rapidly along a radial path and thus the radiation emitted
in the optically thick regions of the flow is reabsorbed locally.
This is due to the very steep radial dependence of the synchrotron
absorption coefficient.

There are several issues we would like to address regarding the
relative importance of thermal and non-thermal electrons in the flow
and the energy exchange between the two populations. First, the
quantity $[\delta_{\rm{nt}}/\delta](r)$ determines as a function of
radius the relative rate of heating of non-thermal electrons compared
to thermal electrons.  Second, the energy exchange between the
populations (heating of the thermal electrons by non-thermal
electrons) proceeds both via the Coulomb term and the absorption of
the synchrotron photons emitted by the power-law populations (the
second term of equation~(\ref{eq:enth}). Therefore the magnitude of
these terms relative to the viscous heating of the thermal electrons
needs to be assessed. Finally, the importance of the thermalization of
the power-law distribution due to the Coulomb term also needs to be
understood.
 
We first estimate the relative magnitude of the terms in
equation~(\ref{eq:ennt}). The ratio of the energy loss of the
power-law electrons due to Coulomb collisions (the last term) to the
energy emitted in synchrotron photons (the second to last term)
determines whether Coulomb collisions or synchrotron self-absorption
is the dominant mechanism by which the non-thermal electrons cool.
The hybrid spectra presented in this paper have been computed under
the assumption that the power-law electrons lose energy primarily
through synchrotron emission, and this needs to be checked for
consistency. For a given electron velocity $\beta$ in the power-law
distribution, the energy exchange rate with the thermal electrons is
given by (Nayakshin \& Melia 1998)
\begin{equation}
  \label{eq:coulomb}
\left( \frac{dE}{dt} \right)_{ee} \simeq \frac{3}{2}  \frac{\ln \Lambda} 
{t_T} \frac{1} {\beta \gamma_{\rm{av}}}, 
\end{equation}
where $\rm{ln} \Lambda \approx 20 $ is the Coulomb logarithm,
$\gamma_{\rm{av}}$ is the average thermal Lorentz factor, $t_T \equiv
(n_e c \sigma_T)^{-1}$ is the Thomson mean-free time and $n_e$ the
electron density.  The rate of energy loss of an electron of Lorentz
factor $\gamma$ emitting synchrotron radiation in a region of magnetic
field B is (Rybicki \& Lightman 1979)
\begin{equation}
  \label{eq:esync}
\left( \frac{E}{t_{1/2}} \right)_{\rm{syn}} \simeq \frac{2 e^4 B^2 \gamma^2} 
{3 m^2 c^3},   
\end{equation}
where $t_{1/2}$ is the time for the electron to lose half its energy.
Using the analytic expressions for the electron density and magnetic
field strength given in $\S 2.1$ and substituting values for the
coefficients and parameters appropriate for Sgr~A$^*$, we find for an
electron of a given $\gamma$:
\begin{equation}
 \label{eq:ratio}
\frac{(dE/dt)_{ee}}{(E/t_{1/2})_{\rm{syn}}} \approx 0.1 r \gamma^{-2}, 
\end{equation} 
where r is the radius in Scwarzschild units as usual. Most of the
emission in the low-frequency shoulder in the spectrum originates from
around $r \simeq 100$ and the radiation is produced by electrons with
$\gamma \simeq 50$ (cf. Fig.~8c and the Appendix). We thus see that
the ratio in equation~(\ref{eq:ratio}) is at most a few percent.  This
shows that Coulomb collisions play a negligible role in the cooling of
non-thermal electrons. It also shows that that the heating of thermal
electrons by Coulomb collisions is unimportant compared to energy
exchange via synchrotron emission and absorption. We therefore neglect
the Coulomb term in the following calculations.

\vbox{ \centerline{ \psfig{file=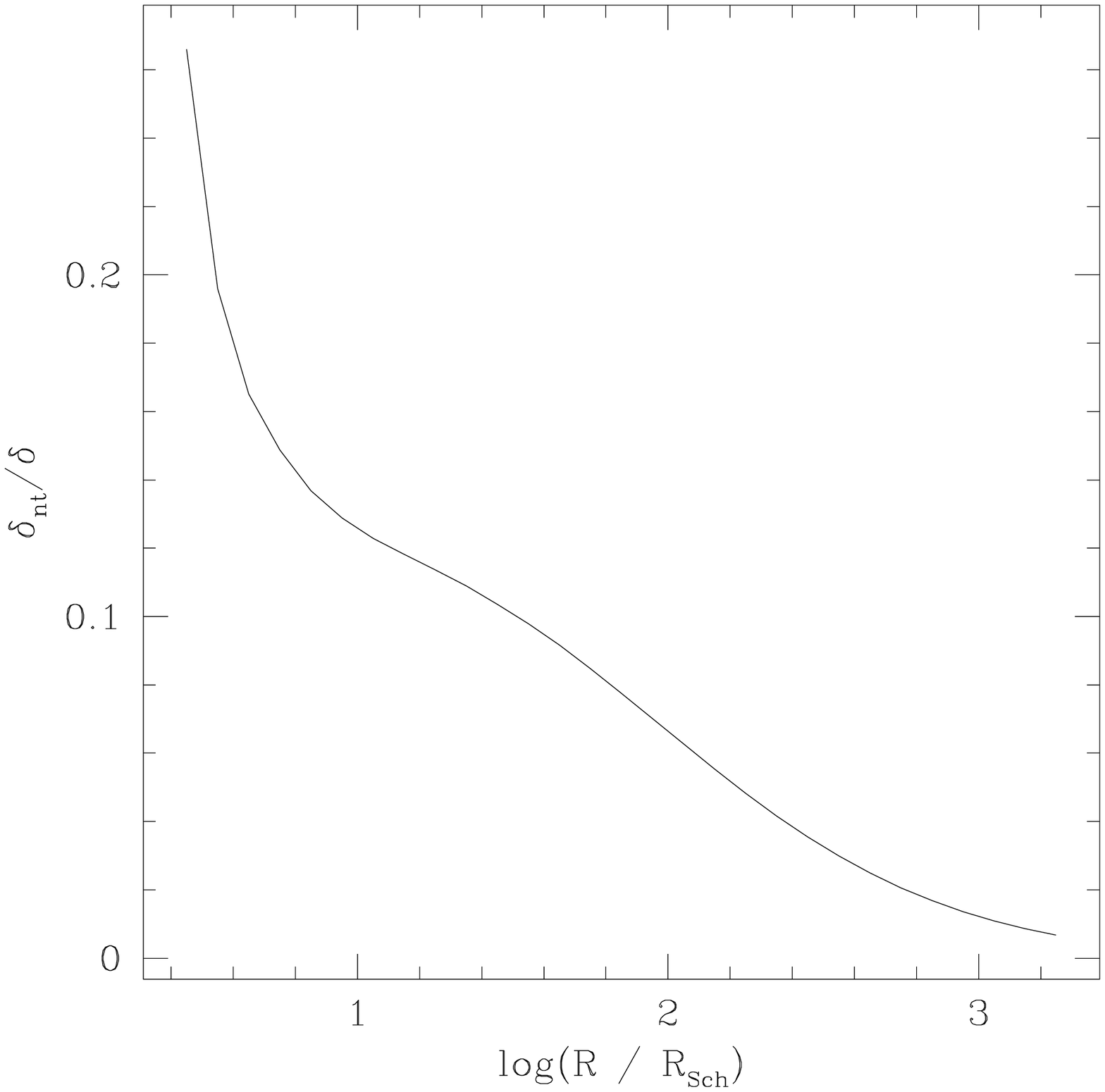,width=9.0truecm} }
\figcaption[]{ \footnotesize The ratio of the non-thermal electron
  heating parameter $\delta_{\rm{nt}}$ to the thermal electron heating
  parameter $\delta$ as a function of radius for a model of Sgr~A$^*$
  with $p=3.5$, $\eta = 0.5\%$.  }}
\vspace*{0.5cm}

The power-law electrons then obey a simple energy balance between the
injected energy through viscous dissipation and the energy loss
through synchrotron emission.  The latter simply is the total
frequency- and angle-integrated synchrotron emissivity of the
non-thermal electrons because this population is optically thin to
synchrotron emission.  Therefore, once $\delta_{\rm{nt}}(r)$ is
specified, it is possible to calculate the parameter $\eta$, which
describes the steady-state energy content of the power-law electrons,
as a function of radius.  Conversely, if we specify $\eta(r)$ as we
did in the previous sections, it is possible to compute the energy
$\delta_{\rm{nt}}$ that would need to be injected into the non-thermal
electrons as a function of radius.  We first study the latter case for
$\eta = $ constant, as this is the assumption we have made in most of
this paper.  Figure 9 shows $\delta_{\rm{nt}} / \delta$ as a function
of radius for one of the hybrid models of Sgr A* with $\eta=0.5\%$,
and $p=3.5$.  The figure demonstrates that the energy input into the
non-thermal electrons never exceeds $20 \%$ of the heating of the
thermal electrons and in fact does not exceed $10\%$ at those radii
that contribute to the low-frequency shoulder in the spectrum $(10R_s
< R < 1000R_s)$.  Thus, at no radius in the flow do the power-law
electrons become energetically more important than, or even comparable
to, the thermal electrons. This may seem like a surprising result
considering that the non-thermal radiation clearly dominates over the
thermal emission in the low-frequency shoulder of the spectrum and in
the high-frequency tail.  However, the synchrotron emissivities of the
thermal and power-law populations are different by many orders of
magnitude at these frequencies. (Optically thin synchrotron emission
from a power-law population peaks at a much higher intensity and at a
lower frequency than the thermal emission for the same energy density
and magnetic field strength.)  Therefore, even when the non-thermal
population has less energy content than the thermal population, and
even when a fraction of the non-thermal radiation is absorbed by the
thermal electrons at frequencies where the flow is optically thick,
the escaping radiation can still be dominated by the non-thermal
emission and the intensity can be much above the blackbody limit.
This is the situation in the low-frequency shoulder of the spectrum.
In the high-frequency tail, the only particles that are energetic
enough to produce the radiation are the non-thermal electrons, and
since the flow is optically thin all the radiation escapes freely.
Incidentally, the fact that the energy input into the non-thermal
electrons does not exceed $20 \%$ of the heating of the thermal
electrons (and even this level is reached only over a small range of
radius close to the black hole) shows that the dynamics of the flow is
not affected by the presence of non-thermal electrons.

We also consider the other mechanism of energy exchange between the
two populations, namely synchrotron self-absorption by thermal
electrons. The second term in equation~(\ref{eq:ennt}) measures the
maximum energy that can be radiated by the non-thermal electrons as a
function of radius. This term also represents an upper limit on the
energy that can be transferred to the thermal population via
self-absorption. This term is bounded from above by $\delta_{\rm{nt}}
\dot{E}_{\rm{visc}}$ according to equation~(\ref{eq:ennt}).  Since
$\delta_{\rm{nt}} / \delta$ never exceeds $20 \%$ as shown in
figure~9, heating of the thermal electrons by absorbing synchrotron
emission from non-thermal electrons can $\it{at~ most}$ introduce a
$20 \%$ correction to the thermal energy equation; the correction is
only a few percent at large radius. This again justifies neglecting
the contribution of the non-thermal electrons to the thermodynamic
properties of the flow.

Finally we note that the calculations in the previous sections
considered a simple parametrization of the non-thermal population via
the quantity $\eta$, which measures the fraction of the electron
energy in steady state that is present in non-thermal electrons. Now,
$\eta$ is a secondary quantity whose value depends on the balance
between non-thermal heating and cooling. It would perhaps be more
useful to parametrize the model by specifying the non-thermal heating
parameter $\delta_{\rm{nt}}$. To this end, we consider models in which
the energy injection rate varies as a power-law in radius,
$\delta_{\rm{nt}} \propto r^{-q}$, and show results for $q$ between 0
and 1. We pick the normalization of $\delta_{\rm{nt}}$ at the inner
edge of the flow such that the resulting $\eta(r)$ is comparable to
the values used in $\S 3$ and is in accord with the results of
Figure~9. In Figure~10a, we show $\eta(r)$ for $p=3.5$ and three
different choices of q: 0, $1/2$, and $3/4$. For none of the three
cases do we see very large variations in the resulting $\eta(r)$
versus radius; we find that $\eta(r)$ rises with radius for $q
\lesssim 1/2$ and falls for larger values of $q$.  Figure~10b shows
the spectra corresponding to the three models of $\delta_{\rm{nt}}$.
We see that the two universal features of a hybrid spectrum, the
low-frequency shoulder and the high-energy tail, are present for all
values of $q$. Thus, super-thermal emission is not an artifact of
specific assumptions or parameter values of our models but is indeed a
robust signature of a hybrid population.  The slope of the
low-frequency shoulder depends very mildly on the particular model of
$\delta(r)$.

\vbox{ \centerline{ \psfig{file=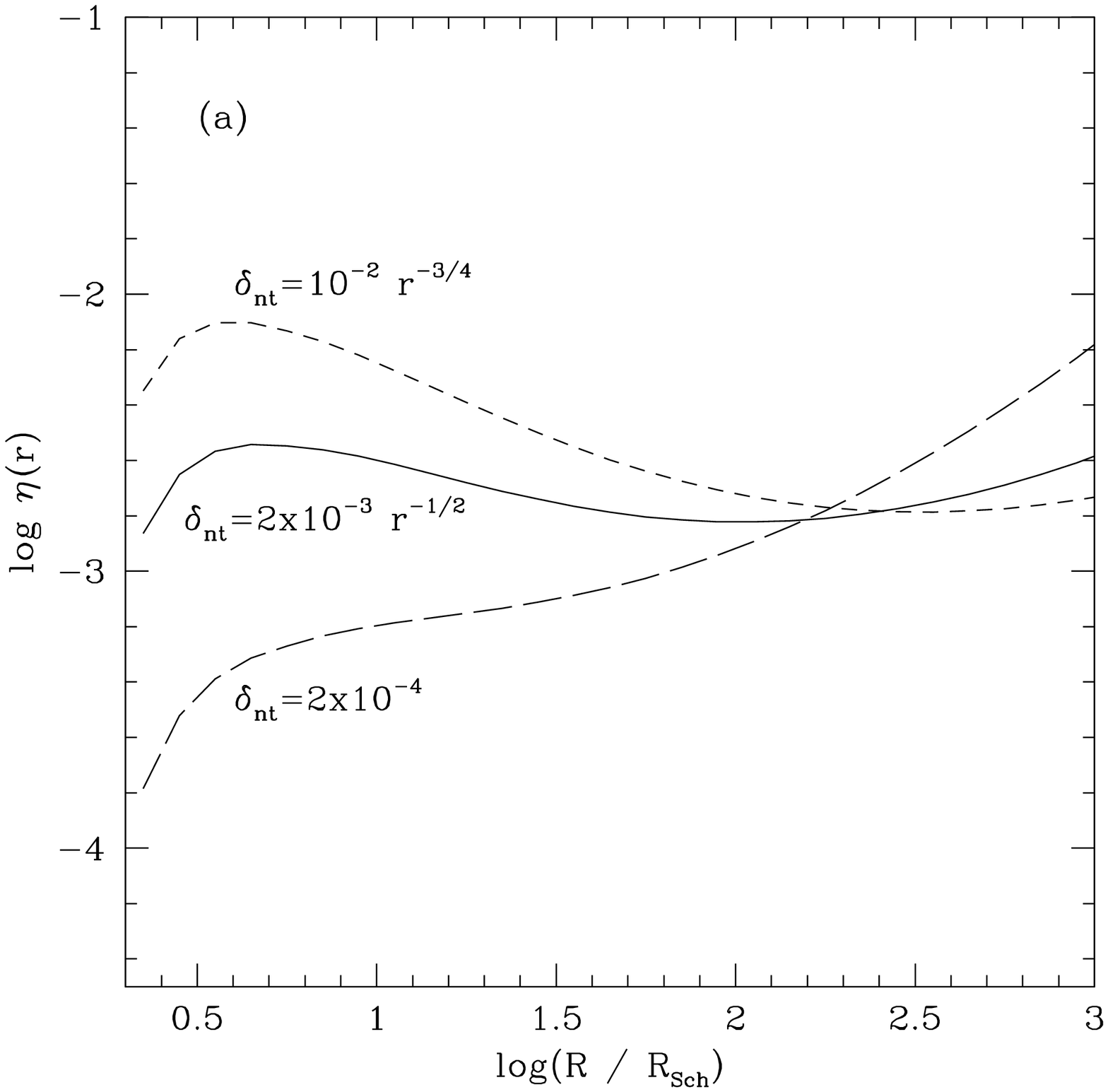,width=7.truecm} 
 \psfig{file=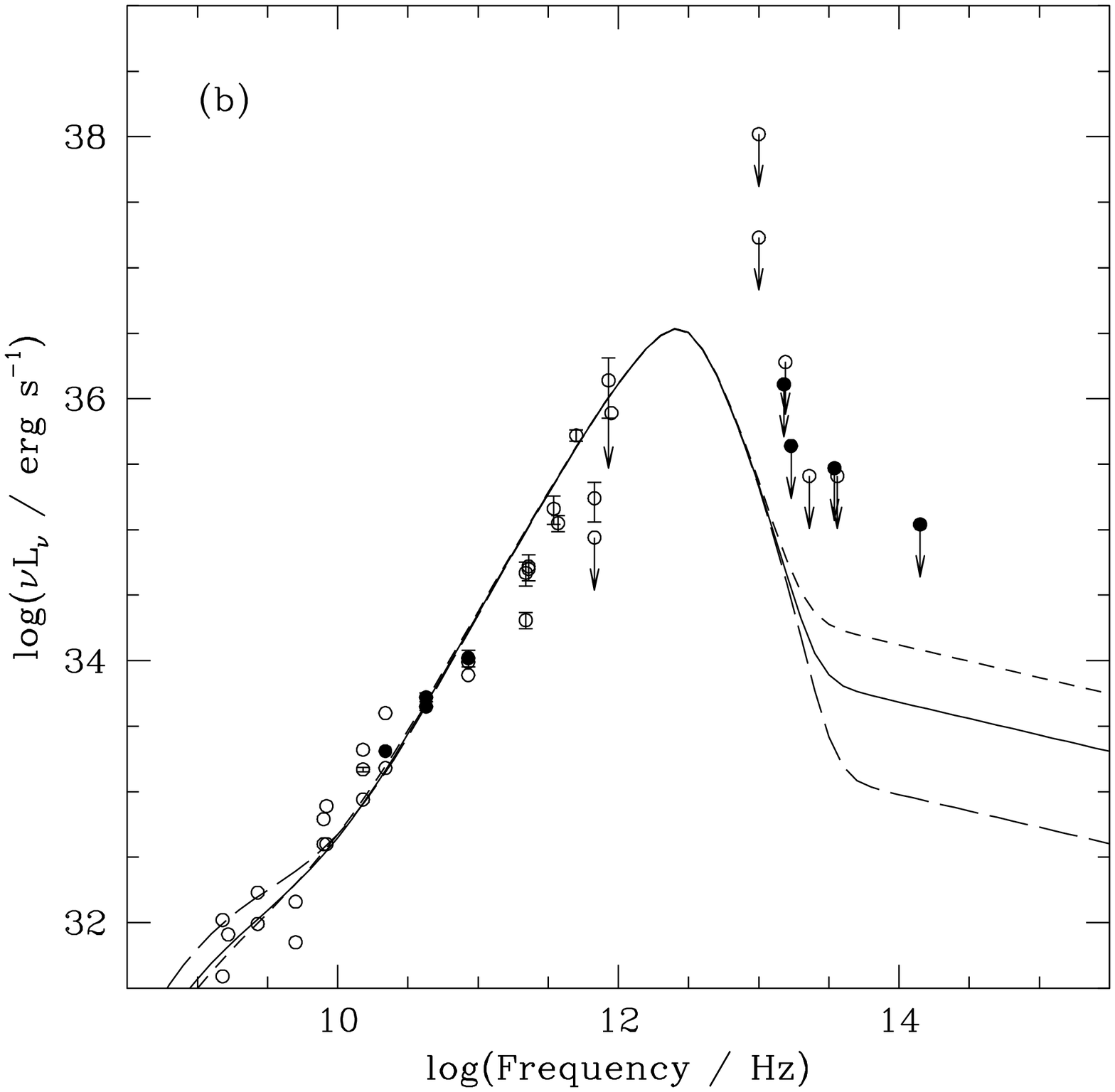,width=7.truecm} }
\figcaption[] { \footnotesize (a) The steady-state non-thermal energy
  content $\eta(r)$ corresponding to three models of the non-thermal
  energy injection rate, $\delta_{\rm{nt}} \propto r^{-q}$, with
  $q=0$, $1/2$, and 1.  All the models have $p=3.5$.  (b) Spectra
  corresponding to the same three models, compared with data on
  Sgr~A$^*$.  }}
\vspace*{0.5cm}

\section{Discussion}

In this paper we considered hot accretion flows around supermassive
black holes, using the ADAF model as a typical example. Such hot flows
are expected to occur at low mass-accretion rates. We assume that a
fraction of the viscous dissipation energy in the accretion flow goes
into accelerating electrons to a non-thermal power-law distribution.
We find that a power-law tail of high energy electrons can be
sustained in such flows; neither Coulomb collisions nor synchrotron
self-absorption is able to thermalize the power-law electrons.
Assuming that there are no other thermalizing mechanisms (e.g.,
collective plasma modes), we calculate the resulting hybrid
thermal/non-thermal spectrum from such a plasma.

The presence of even a population of non-thermal electrons gives rise
to two universal and prominent features in the synchrotron spectrum: a
low-frequency shoulder and a high-frequency tail (Fig.~2).  These
features were identified by Mahadevan (1998) who considered a specific
mechanism (via pion decay) for the production of power-law electrons.
Even if only a small fraction of the total steady-state electron
energy is in the non-thermal power-law component, we find that there
is significant super-thermal emission in the low-frequency shoulder
and the high frequency tail of the spectrum.  Furthermore, each of
these universal features can be associated with a specific range of
Lorentz factors of the emitting electrons. The low-frequency shoulder
is emitted by electrons with Lorentz factors $\log \gamma = 1-2$,
while the high-energy tail is emitted by electrons with $\log \gamma
\gtrsim 2$ (Fig.8).

Since the power-law electrons cause significant emission at low
frequencies, comparing this low-frequency shoulder to data can provide
stringent constraints on the fraction of the electron energy that is
present in a non-thermal population. In the case of Sgr~A$^*$
(Fig.~4), we conclude that, in steady-state, electrons even with
intermediate Lorentz factors ($\gamma$ as low as 10) make up only a
very small fraction of the electron population.  Since Coulomb
collisions and synchrotron self-absorption are ineffective in
thermalizing these flows, this means that either the energy injected
into a non-thermal population, $\delta_{\rm{nt}} \dot{E}_{\rm{visc}}$,
is small as in the calculations described in $\S 5$ or there is some
other efficient thermalization mechanism which plays a role in these
flows. The study on Sgr~A$^*$ needs to be extended to other sources
before we can judge how universal the result is.

Here it is important to emphasize two points regarding the constraints
on non-thermal electrons obtained from the spectrum.

First, for the part of the spectrum below the thermal peak, there is a
mapping between each frequency in the spectrum and the thin shell of
radii which predominantly emits at that frequency. This is due to the
steep radial dependence of the synchrotron absorption that causes a
sharp transition between optically thick and optically thin regions in
the flow. Therefore, emission at each frequency comes primarily from a
specific radius in the flow and can be used to probe the non-thermal
content in a shell that extends from that radius out to a somewhat
larger radius. (Mathematically, radiation emerging at each frequency
comes from the entire range from the radius where the optical depth is
unity out to infinity, but because the total emissivity also drops
rapidly with radius, the biggest contribution to the intensity
integral comes from a relatively narrow shell).  As a result, the
constraints on the non-thermal distribution obtained from each
frequency of the low-frequency shoulder apply only to electrons within
this shell.  Inside the radius corresponding to optical depth unity,
the emission is heavily self-absorbed and the electron distribution
there is inaccessible to observations.

Second, the constraints on the non-thermal energy content which we
have obtained for Sgr~A$^*$ would not be invalidated if the observed
low-frequency shoulder originated in a separate, extended region such
as a jet or an outflow exterior to the accretion flow. In that case,
the observed fluxes may be treated as upper limits to the emission
from the inner accretion flow. This would imply that the non-thermal
emission from the accretion flow is even smaller than in our models
and the resulting constraints on the fraction of non-thermal electrons
would become only tighter.

When the non-thermal electron distribution is a power law, three
parameters of the model and three corresponding pieces of physics can
be extracted by comparing hybrid emission models to spectral data.
The first parameter is $\eta$, or equivalently $\delta_{\rm{nt}}$
(cf.\S5), which measures the fraction of the dissipated energy that
goes into the non-thermal electrons.  The second parameter is $p$, the
slope of the power-law energy distribution (eq.~4).  This can be
uniquely determined by measuring the spectral slope of the
high-frequency tail in the spectrum (provided there is no competing
emission via Compton scattering at these wavelengths, cf. Fig. 1).  The
value of $p$ may provide an indication of the mechanism by which the
electrons in a hot accretion flow are accelerated into a power-law
tail. The third parameter, which is more significant for distributions
with $p<3$, is $\gamma_{\rm{max}}$, the maximum Lorentz factor to
which electrons are accelerated. If this parameter can be determined
by an IR cutoff in the spectrum, it would provide information on the
ratio of the acceleration timescale to the synchrotron cooling
timescale.

Due to the degeneracy of the low-frequency shoulder to the parameters
$\eta$ and $p$ (Fig.~4), this segment of the spectrum by itself is not
sufficient to constrain the two parameters.  However, $\eta$ and $p$
may be determined uniquely if we have information on both the
low-frequency shoulder and the high-frequency tail.  For this to work,
there should be no contamination from a jet or outflow to the observed
radiation in the low-frequency shoulder or from Compton scattering in
the high frequency tail.

We showed in this paper that the presence of a non-thermal population
of electrons also has measurable effects on the shape of the image of
the source, and the size of the accretion flow, as a function of
frequency (Figs.~5-7). Because the frequency dependence of the image
size is strongly affected by the fraction of non-thermal electrons in
the flow, measuring the size of accretion flows at multiple
wavelengths can provide a quantitative constraint on the non-thermal
electron energy content.  Moreover, the brightness temperature of the
source as a whole, as well as the variation of the brightness
temperature across the source, at long wavelengths behave differently
for a hybrid plasma compared to a purely thermal population of
electrons.  Finally, at the frequencies corresponding to the
low-frequency shoulder, we show that there would be limb brightening
of the image.  All these image-related signatures may help to identify
the presence of a non-thermal population of electrons, and the
parameters of the corresponding energy distribution.

\acknowledgements We thank Eliot Quataert, Anthony Aguirre, Tiziana Di
Matteo, and James Cho for many useful discussions and comments on the
manuscript. We also thank S. Nayakshin, the referee, for many useful
comments and suggestions that improved the presentation of the paper.
D. P. was supported by a post-doctoral fellowship of the
Smithsonian Institution. This work was supported in part by NSF Grant
AST 9820686.

\appendix

\section{Analytic Approximations}

In this appendix, we provide analytic formulae for the slope of the
spectrum below and above the thermal peak and the relative importance
of non-thermal electrons as a function of mass, accretion rate, power
law index $p$, and non-thermal energy content $\eta$. For convenience,
we use the self-similar solution of ADAFs to obtain these analytic
estimates.

\subsection{Self-Similar Advection-Dominated Flows}

We begin by presenting the self-similar ADAF solution developed by
Narayan $\&$ Yi (1994, 1995b).  The self-similar solution describes
the local properties of the accretion flow as a function of the black
hole mass $M$, the mass accretion rate $\dot{M}$, the radius $R$, the
viscosity parameter $\alpha$, the ratio of gas pressure to the total
pressure $\beta$, and the fraction of viscously dissipated energy that
is advected inwards $\it{f}$.
 
In terms of the scalings introduced in $\S 2.1$, the height-averaged
electron number density $n_e$, the magnetic field strength B, and the
dimensionless proton temperature $\theta_p \equiv k T_p/m_p c^2$ of
the accretion flow are:
\begin{equation}
  \label{eq:adafsoln1}
  n_e = n_1 m_6^{-1} \dot{m}_{-3} r^{-3/2} ~ {\rm g ~ cm^{-3}}, 
\end{equation} 
\begin{equation} 
  \label{eq:adafsoln2}
  B = b_{1} m_6^{-1/2} \dot{m}_{-3}^{1/2} r^{-5/4} ~ {\rm G}, \\
\end{equation} 
and 
\begin{equation}
  \label{eq:adafsoln3}
  \theta_p = 0.18 \beta r^{-1}, 
\end{equation}
where 
\begin{equation}
  \label{eq:adafcoef1}
  n_1 = 2.0 \times 10^{10} \alpha^{-1} c_{1}^{-1} c_{3}^{-1/2}, \\
\end{equation}
and  
\begin{equation}
  \label{eq:adafcoef2}
  b_{1} = 2.07 \times 10^{4} \alpha^{-1/2} (1 - \beta)^{1/2} c_{1}^{-1/2} 
  c_{3}^{1/4}. 
\end{equation}
The coefficients $c_{1}$ and $c_{3}$ are defined in Narayan $\&$ Yi
(1995b) and are related to the adiabatic index of the gas and the
fraction of advected energy $\it{f}$. For the cases of interest here,
$c_{1} \simeq 0.5$ and $c_{3} \simeq 0.3.$ The remaining parameters
were specified in $\S 2.1$.

\subsection{Fundamental synchrotron quantities in self-similar ADAFs} 

We now use the self-similar solution to obtain expressions for the
cyclotron frequency and the harmonic number for synchrotron emission.
Since we focus our applications to galactic nuclei and very large mass
black holes where the synchrotron frequencies of interest are in the
radio regime, we scale the frequency as $\nu_{10} = \nu / 10^{10}.$
The magnetic field strength is then
\begin{equation}
B = 4.85 \times 10^{4} m_{6}^{-1/2} \dot{m}_{-3}^{1/2} r^{-5/4}
\end{equation}
and thus fundamental cyclotron frequency defined in $\S 2.3$ becomes:
\begin{equation}
\nu_{b} = 1.4 \times 10^{11} m_{6}^{-1/2} \dot{m}_{-3}^{1/2} r^{-5/4}
\end{equation}
The ion temperature retains its virial value throughout the flow and
is given, in dimensionless units $\theta_i = \frac{k T_i}{m_u c^2}$,
by:
\begin{equation}
\theta_{i} = 0.61 \beta c_3 r^{-1} = 0.2 r^{-1}
\end{equation}
when we set the parameter values to those used in the simulations.
The behaviour of the electron temperature on the other hand is more
complicated and depends on the detailed balance of heating and cooling
at each radius.  Qualitatively, electron temperature starts out virial
and equal to the ion temperature at large radii $(r = 10^5)$, but
increases less steeply than the ion temperature at smaller radii due
to cooling by mainly synchrotron emission.  Because it is not possible
to derive the exact electron temperature as a function of radius
analytically, we provide instead numerical fits to the average radial
dependence of electron temperature
\begin{equation}
\theta_e = 1.65 \times r^{-0.6}.
\end{equation}

Since synchrotron absorption coefficient is a steep function of
radius, we can assume that most of the optically thick emission comes
from a narrow range of radii around the radius with optical depth
equal to unity.  Due to the complicated nature of the expression for
the thermal synchrotron absorption which does not allow analytic
integration, we numerically determine the dependence of this $\tau =
1$ radius, $r_{t1}$ on mass, mass accretion rate and frequency. The
best fit gives
\begin{equation}
  \label{eq:tau1}
r_{t1} = 2.5 \times 10^{2} m_{6}^{-1/4} \dot{m}_{-3}^{1/3} \nu_{10}^{-0.6}
\end{equation} 
Substituting $r_{t1}$ into $\nu_{b}$ and $\theta_{e}$ we get
\begin{equation}
\nu_{b} = 1.4 \times 10^{8} m_{6}^{-3/16} \dot{m}_{-3}^{1/12} \nu_{10}^{3/4}
\end{equation} 
and 
\begin{equation}
\theta_{e} = 0.06 \times m_{6}^{0.15} \dot{m}_{-3}^{-0.2} \nu_{10}^{0.36}
\end{equation}
Finally, we compute the harmonic number defined in $\S 2.3$:
\begin{equation}
x_{M} = \frac{2 \nu}{3 \nu_{b} \theta_{e}^{2}}
\end{equation}
\begin{equation}
x_{M} = 1.32 \times 10^{4} m_{6}^{-0.1} \dot{m}_{-3}^{0.3} \nu_{10}^{-0.47}
\end{equation}
which provides all the necessary synchrotron expressions as a function
of black hole mass, mass accretion rate and frequency.

\subsection{Where does the non-thermal emission dominate: Normalization}

We can now derive analytic estimates for the contribution of the
non-thermal emission to the low-frequency shoulder and the
high-frequency tail. In the optically thick shoulder, self-absorption
of the synchrotron emission is done predominantly by the thermal
electrons, and we find that absorption due to non-thermal electrons
are negligible down to a frequency of $\nu \simeq 10^9$ as discussed
above. Thus the source function will take the form
\begin{equation}
S_{\rm{tot}} = \frac{j_{\rm{th}}}{\alpha_{\rm{th}}} + 
\frac{j_{\rm{pl}}}{\alpha_{\rm{th}}}
\end{equation}
where $j_{\rm{th}}$ is synchrotron emissivity per unit volume given in
$\S 2.3$.

For the analytic approximations, we will use a simplified form of the
thermal emissivity. We first take the fully relativistic limit derived
first by Pacholczyk, which is equivalent to setting $a, b$ and $c$
equal to 1 in $M(x_{M})$. We further neglect the second and third
terms of the sum altogether by setting $b$ and $c$ equal to zero as
these terms provide only a very small correction unimportant for the
present purposes.  Thus, in simplified form, the ratio of the two
emissivities is given by
\begin{equation}
\frac{j_{\rm{pl}}}{j_{\rm{th}}} = 
\eta C^\prime_{\rm{pl}} a(\theta) K_2(1/\theta) \theta
(\frac{\nu}{\nu_b})^{-(p+1)/2} x_M^{1/6} \exp(1.9 x_M^{1/3})
\end{equation} 
where
\begin{equation}
 C^\prime_{\rm{pl}} = \frac{\sqrt{3} C_{\rm{pl}}}{4}
\end{equation}
and all other quantities are as defined as before.

The non-thermal emission dominates over the thermal emission in the
spectrum when $j_{\rm{pl}} /j_{\rm{th}} > 1$. This ratio can be easily
computed using the expressions for $x_M, \theta_e$, and $\nu_b$
evaluated at the $\tau = 1$ radius given above.

The power law emission beyond the thermal peak is optically thin and
hence the excess in this region depends sensitively on the power-law
index as expected and is simply given by the optically thin
synchrotron emission from the power-law electrons. The amount of
emission is determined simply by $j_{\rm{pl}}$.

We finally note that, for the small values of $\eta$ considered here,
thermal emission dominates at the thermal peak. The strong dependence
of the position and the normalization of the thermal peak on $m$ and
$\dot{m}$ is discussed by Mahadevan (1997).

\subsection{Spectral Slope}

The shape of the low-frequency shoulder is dominated by the emission
of the power-law population and absorption by the thermal electrons.
In order to determine the spectral slope of this segment, we therefore
first calculate the hybrid source function
\begin{equation}
S_\nu = j_{\rm{pl}} / \alpha_{\rm{th}}
\end{equation}
at the surface of unit optical depth. To convert this into a
luminosity, we multiply the source function by the area of the
$\tau=1$ surface.  If we write the shoulder of the spectrum as $\nu
L_\nu \propto \nu^s$, the spectral slope $s$ is then given by
\begin{equation}
  \label{thickslope}
s = 2.5 - p/8 - 7.2 \dot{m}_{-3}^{0.1} \nu_{10}^{-0.16} + 
6 m_{6}^{-0.15} \dot{m}_{-3}^{0.2} \nu_{10}^{-0.36}
\end{equation}
which gives the canonical value of $\sim 4/3$ for the mass and
mass-accretion rate of Sgr~A$^*$ and scales (weakly) according to the
expression above for other masses and accretion rates. One can
immediately see that the spectral slope depends very weakly on p, only
as $p/8$, thus demonstrating again the independence of the spectral
slope of the low-frequency shoulder on the shape of the electron
energy distribution.

The slope of the optically thin power law emission is a well-known result,
\begin{equation}
   \label{thinslope}
s = -(p-3)/2
\end{equation}
which in $\nu L_{\nu}$ for $p<3$, flat for p=3 and falls off for
$p>3$.  It is independent of the black hole mass and the accretion
rate.

\end{document}